\documentclass[%
 aip,
 amsmath,amssymb,
 reprint,%
]{revtex4-1}

\usepackage{graphicx}
\usepackage{dcolumn}
\usepackage{bm}

\usepackage[utf8]{inputenc}
\usepackage[T1]{fontenc}
\usepackage{mathptmx}
\usepackage{etoolbox}
\usepackage{booktabs} 
\makeatletter
\def\@email#1#2{%
 \endgroup
 \patchcmd{\titleblock@produce}
  {\frontmatter@RRAPformat}
  {\frontmatter@RRAPformat{\produce@RRAP{*#1\href{mailto:#2}{#2}}}\frontmatter@RRAPformat}
  {}{}
}%
\makeatother
\begin{document}

\preprint{AIP/123-QED}

\title{Relativistic Space-Charge-Limited Current in Electrostatic Limit for a Finite-Width Diode with Arbitrary Injection Energy}
\author{Y. L. Liu}
\email{yaoliliu@gs.ncku.edu.tw}
\affiliation{Institute of Space and Plasma Sciences, National Cheng Kung University, Tainan City 70101, Taiwan}

\author{S. W. Choo}
\affiliation{Science, Mathematics and Technology, Singapore University of Technology and Design, Singapore 487372}

\author{L. K. Ang}
\affiliation{Science, Mathematics and Technology, Singapore University of Technology and Design, Singapore 487372}

\date{\today}

\begin{abstract}
We develop a weakly relativistic and electrostatic formulation of space-charge-limited current (SCLC) with arbitrary injection energy, covering the accelerating-diode and drift-space regimes. The one-dimensional (1D) relativistic SCLC model is obtained from a stationary-current condition, which is  validated against the numerical solutions. 
Based on this relativistic 1D SCLC model, we investigate two different approaches to account for the effects of finite emission area for a two-dimensional (2D) planar diode of a finite width of $W$, which is larger than the gap spacing $D$. 
The field-balance (FB) approach evaluates the opposing space-charge fields at the potential minimum, whereas the density-weighted (DW) approach uses the mean absolute distance of electrons from that minimum. 
The two approaches yield different corrections at finite injection energy, even in the non-relativistic limit, with FB model predicting a larger leading-order enhancement. 
DW model reproduces our previously published classical coefficient and the classical drift-space limit, whereas the FB model gives a different result. 
Note both approaches are identical for zero injection velocity.  
Comparisons with 2D electrostatic particle-in-cell (ES-PIC) simulations, it is found that FB and DW model show better agreement for higher injection energy (= gap voltage) at smaller and larger $W/D$ emission area respectively.        
\end{abstract}

\maketitle

\section{Introduction}

Space-charge-limited current (SCLC) describes the maximum steady current density that can be transported through a vacuum diode under the influence of its self-consistent space-charge field~\cite{Zhang2017,Zhang2021}. When the injected current exceeds this limit, electron reflection and virtual-cathode formation may occur, preventing the establishment of a steady, unidirectional electron flow~\cite{Birdsall_1966}. For a one-dimensional (1D) planar vacuum diode with gap spacing $D$ and applied voltage $V_g$, the classical Child--Langmuir (CL) law~\cite{Child_1911,Langmuir_1913} is given by
\begin{equation}
J_{\rm CL}=\frac{4}{9}\varepsilon_0\sqrt{\frac{2e}{m}}\frac{V_g^{3/2}}{D^2},
\label{Eq_CL}
\end{equation}
where $e$ and $m$ denote the magnitude of the electron charge and the electron mass, respectively, and $\varepsilon_0$ is the vacuum permittivity. For electrons injected with a finite initial kinetic energy $eK_V=mv_0^2/2$ in the nonrelativistic regime, Jaff\'e extended the classical CL law to~\cite{Jaffe_1944_2}
\begin{equation}
J_{\rm Jaffe}=J_{\rm CL}\left(\sqrt{1+\epsilon}+\sqrt{\epsilon}\right)^3,
\label{Eq_Jaffe}
\end{equation}
where $\epsilon=K_V/V_g$ is the normalized injection-energy parameter. The limits $\epsilon\ll1$ and $\epsilon\gg1$ correspond to the accelerating-diode and drift-space regimes, respectively. In the latter limit, the limiting current approaches the classical drift-space CL law, which depends on the injection energy rather than the applied voltage~\cite{Birdsall_1966}.

Extensive studies have extended SCLC theory to account for finite emission area~\cite{Luginsland_1996,Lau_2001,LuginslandPOP2002,RokPRL2003,Koh_2005,Liu_2014,Lau2023,Chernin2024}, short-pulse injection~\cite{Valfells_2002,Koh_2006,Chen2011,Zhu_2011,Kaur2022, Yves2026}, quantum effects~\cite{Lau1991,Ang_2003,AngTPS2024,Ang_2007}, higher-dimensional geometries~\cite{Zhu_2013,ZhuPOP2015,Harsha2021POP,Harsha2021_IEEE,Harsha2021POP,Garner2022}, time-dependent injection~\cite{ValfellsPRL2010, Liu2015_POP, Liu2015_PRSTAB,Huang2019,Liu2017,Zhu2022}, over-injection~\cite{Kadish1986,Fazio1989,Leopold2017,Siman2019,Ashmita2026}, finite initial velocity~\cite{Liu1995,Wang2021,Liao2021,Halpern2022,Zhu2023,Zhu2023_2,Breen2024,Halpern2025}, and transitions between injection-limited and space-charge-limited transport~\cite{LauPOP1994,VerboncoeurPOP2006,KohAPL2006,Koh2008,Chernin2020,Gunnarsson2021,Chua2021,Jassem2021,Jin2023}. These investigations have established that the limiting current depends on the injection conditions, emission geometry, and spatial charge distribution in addition to the applied voltage and electrode separation.

For finite emission areas, the multidimensional SCLC enhancement is often approximated by the factorized expression $J_{\rm SCL}(2D)/J_{\rm SCL}(1D)\simeq1+FG$, where $F$ characterizes the longitudinal electron distribution and $G$ represents the transverse geometrical correction~\cite{Koh_2005}. The conventional longitudinal factor is defined by the density-weighted mean position $F=\langle x\rangle/D$. For the classical CL solution with zero injection velocity, $F=1/4$. For a planar sheet emitter of width $W$, the geometrical factor $G=4D/(\pi W)$ gives the familiar finite-width correction~\cite{Lau_2001}
\begin{equation}
\frac{J_{\rm SCL}(2D)}{J_{\rm SCL}(1D)}
\simeq1+\frac{D}{\pi W}.
\label{Eq_CL2d}
\end{equation}
In the classical drift-space regime, the corresponding finite-width correction becomes~\cite{Liu_2014}
\begin{equation}
\frac{J_{\rm SCL}(2D)}{J_{\rm SCL}(1D)}
\simeq1+\frac{7D}{8\pi W}.
\label{Eq_DCL2d}
\end{equation}
The coefficient $7/8$ reflects the longitudinal charge distribution in the drift-space regime and should not be identified directly with the conventional mean-position factor $F$. These results demonstrate that the finite-width enhancement depends on the longitudinal electron distribution even when the transverse emission geometry remains unchanged.

The conventional finite-width corrections are generally developed for sufficiently wide emitting regions, and their applicability to narrow emitters is not guaranteed. Recently, Lau et al.~\cite{Lau2026} investigated the limiting current for narrow emitting regions and low injection energies using differential and integral formulations together with particle-in-cell (PIC) simulations. Their results identified distinct scaling behaviors determined by the relative magnitudes of the normalized emission width and injection energy. These findings emphasize the importance of considering both geometrical and injection conditions when assessing the validity of finite-width approximations. More recently, Koizumi et al.~\cite{Koizumi2026} demonstrated that internal electrostatic staging can modify the limiting-current condition even under fixed terminal boundary conditions, further highlighting the importance of the internal potential structure in space-charge-limited transport.

In our previous work~\cite{Liu2024scl}, we investigated nonrelativistic two-dimensional SCLC with arbitrary injection velocity to connect the accelerating-diode and drift-space regimes. The published finite-width correction recovered the classical limiting coefficients in both regimes. However, re-examination of the original field-balance (FB) derivation reveals inconsistencies in the treatment of the integration direction and inverse-tangent branch. As demonstrated in Appendix~\ref{App_2024_FB_correction}, correcting these inconsistencies yields a different FB coefficient, whereas the previously published coefficient can be recovered independently through the density-weighted (DW) formulation developed in the present work. This distinction indicates that the two approaches employ different physical assumptions and motivates a more systematic examination of their finite-width predictions.

When the injection energy or applied voltage becomes comparable to the electron rest energy, relativistic effects must be incorporated into the longitudinal electron dynamics. The electron velocity, charge density, and position of the potential minimum are then determined by relativistic energy conservation, and the corresponding finite-width corrections cannot generally be described by classical longitudinal factors. Nevertheless, a relativistic treatment of longitudinal motion does not constitute a complete electromagnetic description of multidimensional electron transport. In particular, an electrostatic formulation neglects the self-generated magnetic fields, whose effects may become important at relativistic electron velocities. The present study therefore focuses on electrostatic SCLC, with particular emphasis on the weakly relativistic regime in which this approximation is more readily justified. Predictions at higher relativistic energies are examined with this limitation in mind.

In this work, we develop a relativistic 1D SCLC formulation for arbitrary injection energy and extend it to a planar diode with a finite-width sheet emitter. The 1D limiting current is determined from a stationary-current condition, which also provides the minimum Lorentz factor $\gamma_m$ and the position $x_m$ of the potential minimum. The theoretical results are evaluated against independently obtained numerical boundary-value-problem (BVP) solutions. Using the resulting longitudinal electron distribution, we calculate the relativistic mean-position factor $F$ as a function of the dimensionless gap voltage $U=eV_g/(mc^2)$ and injection parameter $\epsilon=K_V/V_g$, with the relativistic injection energy defined by $eK_V=(\gamma_0-1)mc^2$.

For the finite-width extension, we examine two distinct approximations based on the same relativistic 1D SCLC solution. The FB approach evaluates the longitudinal electric-field balance at the potential minimum, whereas the DW approach introduces the effective longitudinal factor $F_{\rm eff}=\langle|x-x_m|\rangle/D$, representing the density-weighted mean distance from that minimum. The two approaches are compared analytically and evaluated against two-dimensional electrostatic PIC simulations performed using XOOPIC~\cite{Verboncoeur1995}. Particular attention is given to the influence of injection energy, applied voltage, and emitter width, together with the physical assumptions and limitations of both approximations.

The remainder of this paper is organized as follows. Section~II presents the relativistic 1D SCLC formulation and its numerical validation, together with the longitudinal distribution factors. Section~III develops the FB and DW finite-width approximations and compares their predictions with electrostatic PIC simulations. The physical implications and limitations of the models are discussed in Sec.~IV, followed by the conclusions in Sec.~V.

\section{One-Dimensional Relativistic Space-Charge-Limited Current}

\begin{figure}[htbp]
    \centering
    \includegraphics[scale=0.7]{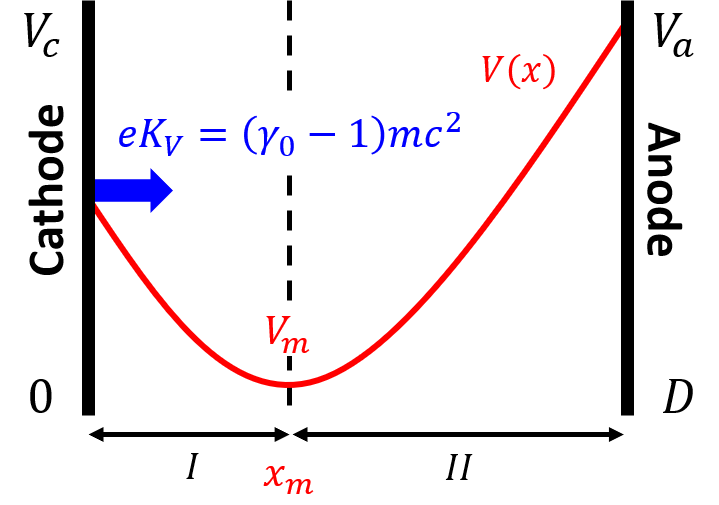}
    \caption{
    Schematic illustration of the electrostatic potential profile in a vacuum diode with electrons injected at an initial velocity $v_0$. The minimum potential and its position are denoted by $V_m$ and $x_m$, respectively. The kinetic voltage $K_V$ is defined by $eK_V=(\gamma_0-1)mc^2$, where $\gamma_0=(1-v_0^2/c^2)^{-1/2}$. In the nonrelativistic limit, this relation reduces to $eK_V=mv_0^2/2$~\cite{Liu2024scl}.
    }
    \label{Fig_1}
\end{figure}

We consider a steady, uniform electron flow injected into a planar vacuum diode with an initial velocity $v_0$, as illustrated in Fig.~\ref{Fig_1}. The cathode and anode are located at $x=0$ and $x=D$, respectively. Conservation of relativistic energy gives
\begin{equation}
(\gamma-1)mc^2=e\left[V(x)+K_V\right],
\label{Eq_energy_conservation}
\end{equation}
where $\gamma=(1-v^2/c^2)^{-1/2}$ is the Lorentz factor, $V(x)$ is the electrostatic potential, and $K_V$ denotes the kinetic voltage associated with the injection energy. The electron velocity is therefore expressed as
\begin{equation}
v=c\frac{\sqrt{\gamma^2-1}}{\gamma}.
\label{Eq_velocity_gamma}
\end{equation}
Combining Poisson's equation with relativistic energy conservation and the steady-current relation yields
\begin{equation}
\frac{d^2\gamma}{dx^2}
=
\frac{eJ}{\varepsilon_0mc^3}
\frac{\gamma}{\sqrt{\gamma^2-1}},
\label{Eq_poisson_gamma}
\end{equation}
where $J>0$ denotes the magnitude of the electron current density. Multiplying Eq.~\eqref{Eq_poisson_gamma} by $2\,d\gamma/dx$ and integrating once gives
\begin{equation}
\left(\frac{d\gamma}{dx}\right)^2
=
\frac{2eJ}{\varepsilon_0mc^3}
\left[\sqrt{\gamma^2-1}-\sqrt{\gamma_m^2-1}\right],
\label{Eq_first_integral}
\end{equation}
where $\gamma_m=\gamma(x_m)$ is the minimum Lorentz factor and $(d\gamma/dx)_{x=x_m}=0$.

The diode is divided into two regions by the position of the potential minimum. Region I extends from the cathode to $x_m$, with $0<x<x_m$, whereas region II extends from $x_m$ to the anode, with $x_m<x<D$. The longitudinal gradient of the Lorentz factor is consequently
\begin{equation}
\frac{d\gamma}{dx}
=
\begin{cases}
-\displaystyle
\left(\frac{2eJ}{\varepsilon_0mc^3}\right)^{1/2}
\sqrt{\sqrt{\gamma^2-1}-\sqrt{\gamma_m^2-1}},
& 0<x<x_m, \\[3mm]
+\displaystyle
\left(\frac{2eJ}{\varepsilon_0mc^3}\right)^{1/2}
\sqrt{\sqrt{\gamma^2-1}-\sqrt{\gamma_m^2-1}},
& x_m<x<D.
\end{cases}
\label{Eq_gamma_gradient}
\end{equation}
The boundary conditions are
\begin{equation}
\gamma(0)=\gamma_c,\qquad
\gamma(x_m)=\gamma_m,\qquad
\gamma(D)=\gamma_a,
\label{Eq_boundary_conditions}
\end{equation}
where
\begin{equation}
\gamma_c=1+\frac{e(V_c+K_V)}{mc^2},\qquad
\gamma_a=1+\frac{e(V_a+K_V)}{mc^2}.
\label{Eq_gamma_boundaries}
\end{equation}

Integrating Eq.~\eqref{Eq_gamma_gradient} over region I yields
\begin{equation}
g_I(\gamma_c,\gamma_m)
\equiv
\int_{\gamma_m}^{\gamma_c}
\frac{d\gamma}
{\sqrt{\sqrt{\gamma^2-1}-\sqrt{\gamma_m^2-1}}}
=
\left(\frac{2eJ}{\varepsilon_0mc^3}\right)^{1/2}x_m.
\label{Eq_g_I}
\end{equation}
Similarly, integration over region II gives
\begin{equation}
g_{II}(\gamma_a,\gamma_m)
\equiv
\int_{\gamma_m}^{\gamma_a}
\frac{d\gamma}
{\sqrt{\sqrt{\gamma^2-1}-\sqrt{\gamma_m^2-1}}}
=
\left(\frac{2eJ}{\varepsilon_0mc^3}\right)^{1/2}(D-x_m).
\label{Eq_g_II}
\end{equation}
Adding Eqs.~\eqref{Eq_g_I} and \eqref{Eq_g_II}, we obtain
\begin{equation}
g_I(\gamma_c,\gamma_m)+g_{II}(\gamma_a,\gamma_m)
=
\left(\frac{2eJ}{\varepsilon_0mc^3}\right)^{1/2}D.
\label{Eq_g_total}
\end{equation}
The relativistic one-dimensional SCLC density can thus be expressed as
\begin{equation}
J_{\mathrm{SCL}}
=
\frac{\varepsilon_0mc^3}{2eD^2}
\left[g_I(\gamma_c,\gamma_m)+g_{II}(\gamma_a,\gamma_m)\right]^2.
\label{Eq_JSCL}
\end{equation}
The normalized position of the potential minimum follows directly from Eqs.~\eqref{Eq_g_I} and \eqref{Eq_g_total}, giving
\begin{equation}
\frac{x_m}{D}
=
\frac{g_I(\gamma_c,\gamma_m)}
{g_I(\gamma_c,\gamma_m)+g_{II}(\gamma_a,\gamma_m)}
=
\left[1+\frac{g_{II}(\gamma_a,\gamma_m)}
{g_I(\gamma_c,\gamma_m)}\right]^{-1}.
\label{Eq_xm}
\end{equation}

In addition to the limiting current density and the position of the potential minimum, the electron transit time provides a useful characterization of the stationary SCLC state. The relativistic transit time $T_{\mathrm{SCL}}$ is defined as the time required for an electron to traverse the diode gap under steady space-charge-limited conditions. Integrating Poisson's equation with the prescribed electrode potentials, the electric-field magnitudes at the anode and cathode can be expressed as
\begin{equation}
E_a
=
\frac{V_a-V_c}{D}
+
\frac{J_{\mathrm{SCL}}}{\varepsilon_0D}
\int_0^D\frac{x}{v(x)}\,dx,
\label{Eq_Ea_gauss}
\end{equation}
and
\begin{equation}
E_c
=
-\frac{V_a-V_c}{D}
+
\frac{J_{\mathrm{SCL}}}{\varepsilon_0D}
\int_0^D\frac{D-x}{v(x)}\,dx,
\label{Eq_Ec_gauss}
\end{equation}
respectively. Adding these expressions and using $T_{\mathrm{SCL}}=\int_0^D dx/v(x)$ gives
\begin{equation}
E_a+E_c
=
\frac{J_{\mathrm{SCL}}T_{\mathrm{SCL}}}{\varepsilon_0}.
\label{Eq_Ea_Ec_sum}
\end{equation}
This relation holds for an arbitrary longitudinal charge distribution within the stationary one-dimensional model.

Alternatively, from Eq.~\eqref{Eq_first_integral}, the electric-field magnitudes at the electrodes are
\begin{equation}
E_a
=
\frac{mc^2}{e}
\left(\frac{2eJ_{\mathrm{SCL}}}{\varepsilon_0mc^3}\right)^{1/2}
\sqrt{\sqrt{\gamma_a^2-1}-\sqrt{\gamma_m^2-1}},
\label{Eq_Ea}
\end{equation}
and
\begin{equation}
E_c
=
\frac{mc^2}{e}
\left(\frac{2eJ_{\mathrm{SCL}}}{\varepsilon_0mc^3}\right)^{1/2}
\sqrt{\sqrt{\gamma_c^2-1}-\sqrt{\gamma_m^2-1}}.
\label{Eq_Ec}
\end{equation}
Substituting Eqs.~\eqref{Eq_Ea} and \eqref{Eq_Ec} into Eq.~\eqref{Eq_Ea_Ec_sum}, we obtain
\begin{equation}
\begin{aligned}
T_{\mathrm{SCL}}
= 
&\left(\frac{2\varepsilon_0mc}{eJ_{\mathrm{SCL}}}\right)^{1/2}\times \\
&\left[
\sqrt{\sqrt{\gamma_c^2-1}-\sqrt{\gamma_m^2-1}}
+
\sqrt{\sqrt{\gamma_a^2-1}-\sqrt{\gamma_m^2-1}}
\right].
\label{Eq_TSCL}
\end{aligned}
\end{equation}

Equations~\eqref{Eq_JSCL}, \eqref{Eq_xm}, and \eqref{Eq_TSCL} determine the relativistic SCLC density, the position of the potential minimum, and the electron transit time in terms of the minimum Lorentz factor $\gamma_m$. The longitudinal electron distribution is also determined once $\gamma_m$ is specified. We therefore determine $\gamma_m$ from the stationary-current condition before evaluating the relativistic mean-position factor.

\subsection{Minimum Lorentz Factor}

The minimum Lorentz factor $\gamma_m$ is required to determine $J_{\mathrm{SCL}}$, $x_m$, and $T_{\mathrm{SCL}}$ through Eqs.~\eqref{Eq_JSCL}, \eqref{Eq_xm}, and \eqref{Eq_TSCL}, respectively. In the classical regime, the minimum potential can be determined analytically by imposing the stationary-current condition $\partial J_{\mathrm{SCL}}/\partial V_m=0$~\cite{Liu2024scl}. We extend this approach to the relativistic regime by requiring
\begin{equation}
\frac{\partial J_{\mathrm{SCL}}}{\partial\gamma_m}=0.
\label{Eq_stationary_gamma_m}
\end{equation}
Since the prefactor in Eq.~\eqref{Eq_JSCL} is independent of $\gamma_m$, the stationary condition reduces to
\begin{equation}
\frac{\partial g_I}{\partial\gamma_m}
+
\frac{\partial g_{II}}{\partial\gamma_m}=0.
\label{Eq_stationary_g}
\end{equation}

Direct differentiation of Eqs.~\eqref{Eq_g_I} and \eqref{Eq_g_II} is complicated by the integrable singularities at their lower integration limits. To facilitate differentiation, we introduce the normalized relativistic momentum $q=\sqrt{\gamma^2-1}$, with $q_m=\sqrt{\gamma_m^2-1}$, $q_c=\sqrt{\gamma_c^2-1}$, and $q_a=\sqrt{\gamma_a^2-1}$. For either region, we denote the corresponding boundary momentum by $q_j$, where $q_I=q_c$ and $q_{II}=q_a$, and introduce the variable $s=q-q_m$. Using $\gamma=\sqrt{1+(q_m+s)^2}$ and $d\gamma=(q_m+s)[1+(q_m+s)^2]^{-1/2}ds$, both integrals can be expressed in the common form
\begin{equation}
g_j(q_j;q_m)
=
\int_0^{q_j-q_m}
\frac{q_m+s}
{\sqrt{s}\sqrt{1+(q_m+s)^2}}\,ds,
\qquad j=I,II.
\label{Eq_gj_s}
\end{equation}
Applying the Leibniz integral rule to Eq.~\eqref{Eq_gj_s} gives
\begin{equation}
\frac{\partial g_j}{\partial q_m}
=
-\frac{q_j}{\sqrt{1+q_j^2}\sqrt{q_j-q_m}}
+
\int_0^{q_j-q_m}
\frac{ds}{\sqrt{s}\left[1+(q_m+s)^2\right]^{3/2}}.
\label{Eq_dgj_dqm}
\end{equation}

For an interior solution with $\gamma_m>1$, the corresponding momentum satisfies $q_m>0$. Since $dq_m/d\gamma_m=\gamma_m/q_m>0$, the transformation between $\gamma_m$ and $q_m$ is monotonic. To establish the equivalence of the stationary conditions with respect to these two variables, we define
\begin{equation}
G(q_m)=g_I(q_c;q_m)+g_{II}(q_a;q_m).
\label{Eq_G_definition}
\end{equation}
Equation~\eqref{Eq_JSCL} can then be written as
\begin{equation}
J_{\mathrm{SCL}}
=
\frac{\varepsilon_0mc^3}{2eD^2}G^2(q_m).
\label{Eq_JSCL_G}
\end{equation}
Differentiating with respect to $\gamma_m$ yields
\begin{equation}
\frac{\partial J_{\mathrm{SCL}}}{\partial\gamma_m}
=
\frac{\varepsilon_0mc^3}{eD^2}
G(q_m)\frac{\partial G}{\partial q_m}\frac{dq_m}{d\gamma_m}.
\label{Eq_dJSCL_dgammam}
\end{equation}
For a physically admissible interior solution, the prefactor, $G(q_m)$, and $dq_m/d\gamma_m$ are all positive. Therefore, the stationary condition $\partial J_{\mathrm{SCL}}/\partial\gamma_m=0$ is equivalent to
\begin{equation}
\frac{\partial}{\partial q_m}
\left[g_I(q_c;q_m)+g_{II}(q_a;q_m)\right]=0,
\label{Eq_stationary_qm}
\end{equation}
where the boundary momenta $q_c$ and $q_a$ are held fixed during differentiation.

The boundary case $\gamma_m=1$, corresponding to $q_m=0$, requires separate consideration because $dq_m/d\gamma_m$ diverges in this limit. The equivalence established above therefore applies directly only to interior solutions with $q_m>0$.

Substituting Eq.~\eqref{Eq_dgj_dqm} into Eq.~\eqref{Eq_stationary_qm}, we obtain
\begin{align}
&
\int_0^{q_a-q_m}
\frac{ds}{\sqrt{s}\left[1+(q_m+s)^2\right]^{3/2}}
+
\int_0^{q_c-q_m}
\frac{ds}{\sqrt{s}\left[1+(q_m+s)^2\right]^{3/2}}
\nonumber\\
&\qquad =
\frac{q_a}{\sqrt{1+q_a^2}\sqrt{q_a-q_m}}
+
\frac{q_c}{\sqrt{1+q_c^2}\sqrt{q_c-q_m}}.
\label{Eq_qm_implicit}
\end{align}
Equation~\eqref{Eq_qm_implicit} implicitly determines the normalized momentum at the potential minimum. Once the physically admissible root $q_m$ is obtained, the minimum Lorentz factor follows from
\begin{equation}
\gamma_m=\sqrt{1+q_m^2}.
\label{Eq_gamma_m}
\end{equation}
The admissible range is $1\leq\gamma_m<\min(\gamma_a,\gamma_c)$, or equivalently $0\leq q_m<\min(q_a,q_c)$.

Unlike the corresponding classical formulation~\cite{Liu2024scl}, Eq.~\eqref{Eq_qm_implicit} does not yield a simple closed-form expression for $\gamma_m$. The minimum Lorentz factor is therefore determined by numerically solving this scalar nonlinear equation within the physically admissible interval, with the boundary case treated separately when necessary.

\subsection{Relativistic Mean-Position Factor}

We characterize the longitudinal space-charge distribution using the mean-position factor $F$. Introducing the normalized coordinate $\xi=x/D$, with $0\leq\xi\leq1$, the density-weighted average of a function $f(x)$ is defined as $\langle f(x)\rangle=\int_0^D f(x)n(x)\,dx/\int_0^D n(x)\,dx$. The mean-position factor is then given by $F=\langle x\rangle/D$, or equivalently,
\begin{equation}
F
\equiv
\frac{\displaystyle\int_0^1 \xi n(\xi)\,d\xi}
{\displaystyle\int_0^1 n(\xi)\,d\xi}.
\label{Eq_F_definition}
\end{equation}
This dimensionless factor characterizes the longitudinal charge distribution independently of the transverse geometry within the one-dimensional model.

For a steady one-dimensional electron flow, current conservation requires $J=en(x)v(x)$. Substituting $n(x)=J/[ev(x)]$ and using Eq.~\eqref{Eq_velocity_gamma}, the mean-position factor becomes
\begin{equation}
F=
\frac{\displaystyle\int_0^1
\xi\frac{\gamma(\xi)}{\sqrt{\gamma^2(\xi)-1}}\,d\xi}
{\displaystyle\int_0^1
\frac{\gamma(\xi)}{\sqrt{\gamma^2(\xi)-1}}\,d\xi}.
\label{Eq_F_gamma_xi}
\end{equation}
Equation~\eqref{Eq_F_gamma_xi} allows $F$ to be evaluated directly from the stationary Lorentz-factor profile $\gamma(\xi)$.

Alternatively, $F$ can be determined from the analytical one-dimensional SCLC solution. We introduce the partial integral
\begin{equation}
g(\gamma;\gamma_m)
\equiv
\int_{\gamma_m}^{\gamma}
\frac{d\gamma'}
{\sqrt{\sqrt{\gamma'^2-1}-\sqrt{\gamma_m^2-1}}}.
\label{Eq_g_partial}
\end{equation}
Accordingly, $g_I(\gamma_c,\gamma_m)=g(\gamma_c;\gamma_m)$ and $g_{II}(\gamma_a,\gamma_m)=g(\gamma_a;\gamma_m)$. Using Eqs.~\eqref{Eq_g_I}--\eqref{Eq_g_total}, the normalized longitudinal positions in regions I and II are expressed as
\begin{equation}
\xi_I(\gamma)
=
\frac{g_I(\gamma_c,\gamma_m)-g(\gamma;\gamma_m)}
{g_I(\gamma_c,\gamma_m)+g_{II}(\gamma_a,\gamma_m)},
\qquad
\gamma_m\leq\gamma\leq\gamma_c,
\label{Eq_xi_I_F}
\end{equation}
and
\begin{equation}
\xi_{II}(\gamma)
=
\frac{g_I(\gamma_c,\gamma_m)+g(\gamma;\gamma_m)}
{g_I(\gamma_c,\gamma_m)+g_{II}(\gamma_a,\gamma_m)},
\qquad
\gamma_m\leq\gamma\leq\gamma_a.
\label{Eq_xi_II_F}
\end{equation}

From Eqs.~\eqref{Eq_first_integral} and \eqref{Eq_velocity_gamma}, the differential transit time can be written as
\begin{equation}
\frac{|dx|}{v}
=
\frac{1}{c}
\left(\frac{\varepsilon_0mc^3}{2eJ}\right)^{1/2}
h(\gamma;\gamma_m)\,d\gamma,
\label{Eq_dx_over_v_F}
\end{equation}
where
\begin{equation}
h(\gamma;\gamma_m)
\equiv
\frac{\gamma}
{\sqrt{\gamma^2-1}
\sqrt{\sqrt{\gamma^2-1}-\sqrt{\gamma_m^2-1}}}.
\label{Eq_h_F}
\end{equation}
The common prefactor cancels between the numerator and denominator of Eq.~\eqref{Eq_F_gamma_xi}. The relativistic mean-position factor can therefore be expressed entirely in terms of the analytical one-dimensional solution as
\begin{equation}
F
=
\frac{
\displaystyle\int_{\gamma_m}^{\gamma_c}
\xi_I(\gamma)h(\gamma;\gamma_m)\,d\gamma
+
\displaystyle\int_{\gamma_m}^{\gamma_a}
\xi_{II}(\gamma)h(\gamma;\gamma_m)\,d\gamma
}{
\displaystyle\int_{\gamma_m}^{\gamma_c}
h(\gamma;\gamma_m)\,d\gamma
+
\displaystyle\int_{\gamma_m}^{\gamma_a}
h(\gamma;\gamma_m)\,d\gamma
}.
\label{Eq_F_general_rel}
\end{equation}
Once $\gamma_m$ is determined from the stationary-current condition, Eq.~\eqref{Eq_F_general_rel} provides the relativistic mean-position factor for arbitrary injection energy and applied voltage within the one-dimensional formulation.

For the present parameterization, the cathode potential is chosen as the reference, $V_c=0$, so that $V_g=V_a-V_c=V_a$. The dimensionless gap voltage and injection parameter are defined as $U=eV_g/(mc^2)$ and $\epsilon=K_V/V_g$, respectively. The corresponding boundary Lorentz factors are $\gamma_c=1+\epsilon U$ and $\gamma_a=1+(1+\epsilon)U$. Consequently, the mean-position factor depends on these two dimensionless parameters,
\begin{equation}
F=F(U,\epsilon).
\label{Eq_F_U_epsilon}
\end{equation}

\begin{figure}[htbp]
    \centering
    \includegraphics[scale=0.55]{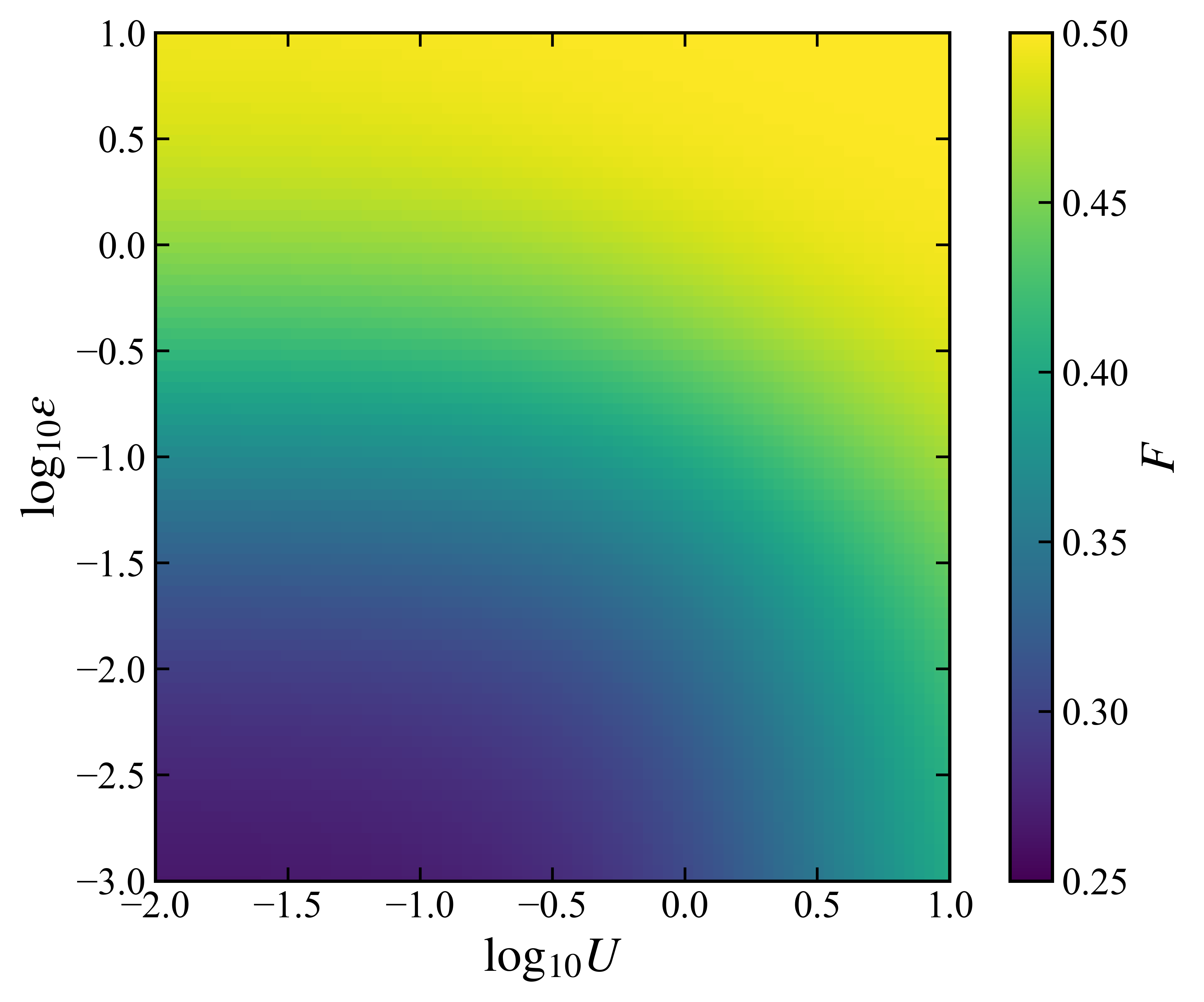}
    \caption{
    Relativistic mean-position factor $F$ as a function of the dimensionless gap voltage $U=eV_g/(mc^2)$ and injection parameter $\epsilon=K_V/V_g$. The classical Child--Langmuir limit and the uniform-density limit correspond to $F=1/4$ and $F=1/2$, respectively.
    }
    \label{Fig_F}
\end{figure}

Figure~\ref{Fig_F} presents the relativistic mean-position factor over the parameter ranges $10^{-2}\leq U\leq10$ and $10^{-3}\leq\epsilon\leq10$. The results show that $F$ increases with both the applied voltage and injection energy, approaching $1/2$ as the longitudinal electron density becomes increasingly uniform. In the zero-injection and weakly relativistic limit, the previously derived expression~\cite{Koh_2005} is recovered,
\begin{equation}
F\simeq\frac{1}{4}\left(1+\frac{3U}{28}\right),
\label{Eq_F_weak_rel}
\end{equation}
which reduces to the classical Child--Langmuir value $F=1/4$ as $U\to0$.

\subsection{Numerical Validation}

We assess the analytical relativistic SCLC formulation through independent numerical solutions of the one-dimensional boundary-value problem (BVP). Introducing the normalized coordinate $\xi=x/D$, Eq.~\eqref{Eq_poisson_gamma} becomes
\begin{equation}
\frac{d^2\gamma}{d\xi^2}
=
\Lambda\frac{\gamma}{\sqrt{\gamma^2-1}},
\label{Eq_normalized_poisson}
\end{equation}
where
\begin{equation}
\Lambda=\frac{eJD^2}{\varepsilon_0mc^3}
\label{Eq_Lambda_definition}
\end{equation}
is the dimensionless current-density parameter. The corresponding boundary conditions are
\begin{equation}
\gamma(0)=\gamma_c,\qquad \gamma(1)=\gamma_a.
\label{Eq_numerical_boundary_conditions}
\end{equation}

The numerical limiting current is estimated using a continuation procedure. Starting from $J=0$, the current density is gradually increased, with each converged solution serving as the initial guess for the subsequent BVP calculation. This procedure follows the stationary solution branch toward the limiting-current condition. As the current approaches the SCLC limit, the solution branch approaches a turning point characterized by $\partial J/\partial\gamma_m=0$. The BVP consequently becomes increasingly ill-conditioned, making numerical convergence sensitive to the prescribed current density and initial guess.

To estimate the limiting current, we identify the largest current density $J_{\mathrm{low}}$ for which a converged and physically admissible solution is obtained, together with a higher current density $J_{\mathrm{high}}$ for which no valid solution is found by the numerical solver. The interval is iteratively refined until
\begin{equation}
\frac{J_{\mathrm{high}}-J_{\mathrm{low}}}{J_{\mathrm{low}}}<\delta_J,
\label{Eq_current_convergence}
\end{equation}
where $\delta_J=2\times10^{-4}$ is the prescribed relative tolerance for the limiting-current search. The resulting estimate, denoted by $J_{\mathrm{SCL}}^{\mathrm{BVP}}$, is taken as the largest current density for which a converged and physically admissible BVP solution is obtained. Since numerical nonconvergence does not necessarily imply the absence of a stationary solution, this procedure provides a numerical estimate of the limiting current rather than a rigorous determination of the critical point.

For each numerical solution at the estimated limiting current, the minimum Lorentz factor is determined from
\begin{equation}
\gamma_m^{\mathrm{BVP}}
=
\min_{0\leq\xi\leq1}\gamma(\xi),
\label{Eq_gamma_m_BVP}
\end{equation}
and the corresponding position of the potential minimum is obtained from
\begin{equation}
\left.\frac{d\gamma}{d\xi}\right|_{\xi=\xi_m}=0,
\qquad
\frac{x_m^{\mathrm{BVP}}}{D}=\xi_m.
\label{Eq_x_m_BVP}
\end{equation}

\begin{figure}[htbp]
    \centering
    \includegraphics[width=0.5\textwidth]{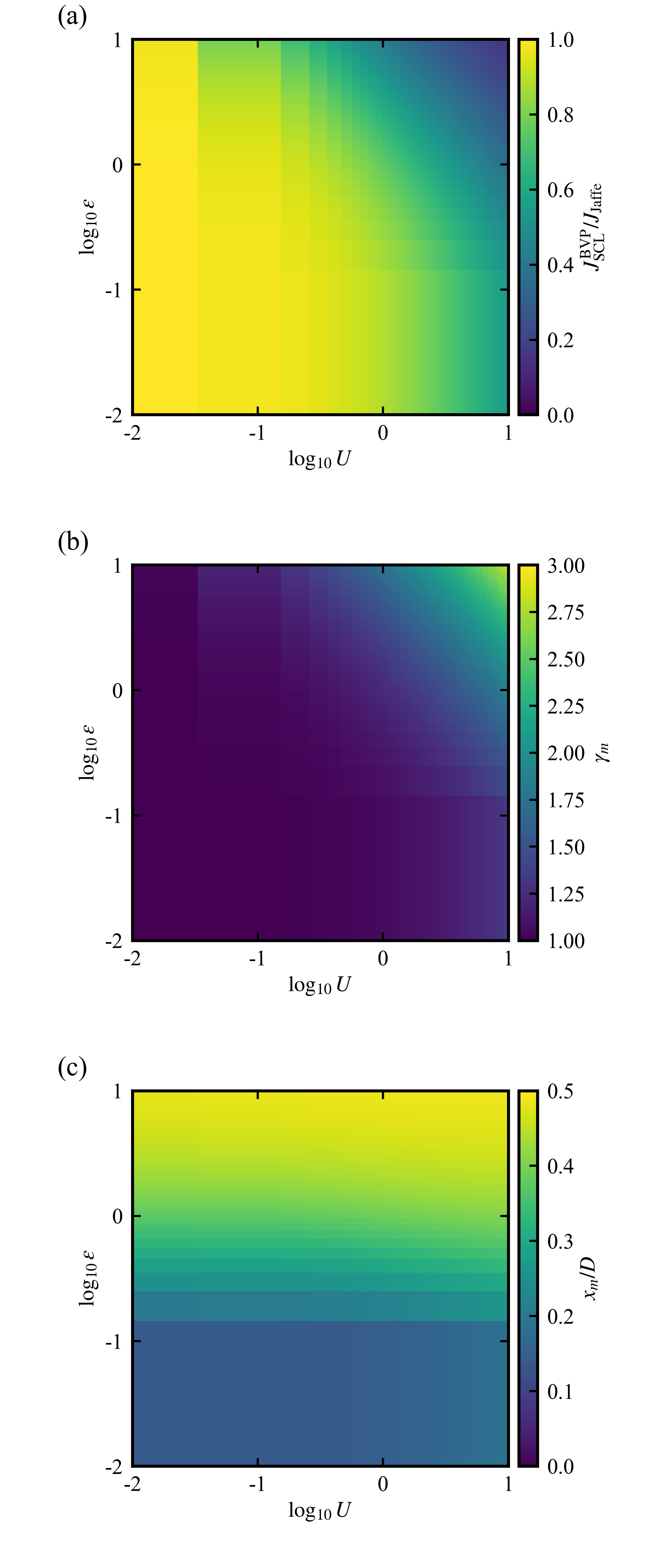}
    \caption{
    Two-dimensional maps of the relativistic SCLC characteristics obtained from the BVP solutions as functions of the dimensionless gap voltage $U=eV_g/(mc^2)$ and injection parameter $\epsilon=K_V/V_g$.
    (a) SCLC current density normalized to the Jaff\'e current density, $J_{\mathrm{SCL}}^{\mathrm{BVP}}/J_{\mathrm{Jaffe}}$.
    (b) Minimum Lorentz factor $\gamma_m^{\mathrm{BVP}}$.
    (c) Normalized position of the potential minimum, $x_m^{\mathrm{BVP}}/D$.
    }
    \label{Fig_map}
\end{figure}

A total of 10,000 numerical limiting-current estimates were obtained for a diode gap of $D=1~\mathrm{cm}$. The parameter scan covers the dimensionless gap voltage $U$ and injection parameter $\epsilon$ defined in Sec.~II, with
\begin{equation}
0.01\leq U\leq10,\qquad
0.001\leq\epsilon\leq10,
\label{Eq_numerical_scan_range}
\end{equation}
corresponding to applied voltages of approximately $5.11~\mathrm{kV}\leq V_g\leq5.11~\mathrm{MV}$. Figure~\ref{Fig_map} presents the normalized limiting-current density, minimum Lorentz factor, and position of the potential minimum over the investigated parameter space.

We next compare the numerical BVP results with the theoretical predictions derived in Sec.~II A. For each combination of $\gamma_a$ and $\gamma_c$, the theoretical normalized momentum at the potential minimum is determined by numerically solving Eq.~\eqref{Eq_qm_implicit} within the physically admissible interval $0\leq q_m<\min(q_a,q_c)$. The resulting solution, denoted by $q_m^{\mathrm{theory}}$, gives the theoretical minimum Lorentz factor through Eq.~\eqref{Eq_gamma_m},
\begin{equation}
\gamma_m^{\mathrm{theory}}
=
\sqrt{1+\left(q_m^{\mathrm{theory}}\right)^2}.
\label{Eq_gamma_m_theory_from_qm}
\end{equation}
The theoretical limiting current $J_{\mathrm{SCL}}^{\mathrm{theory}}$ and the corresponding minimum Lorentz factor $\gamma_m^{\mathrm{theory}}$ are then compared with the independently obtained BVP results.

The relative differences between the theoretical predictions and the BVP results are defined as $\delta_J^{\mathrm{theory}}=(J_{\mathrm{SCL}}^{\mathrm{theory}}-J_{\mathrm{SCL}}^{\mathrm{BVP}})/J_{\mathrm{SCL}}^{\mathrm{BVP}}\times100\%$ for the limiting current density and $\delta_{\gamma_m}^{\mathrm{theory}}=(\gamma_m^{\mathrm{theory}}-\gamma_m^{\mathrm{BVP}})/\gamma_m^{\mathrm{BVP}}\times100\%$ for the minimum Lorentz factor. The theoretical minimum Lorentz factors span the range $1.00001\leq\gamma_m^{\mathrm{theory}}\leq2.72361$. 
Averaging over all 10,000 parameter combinations gives mean absolute relative differences of $\langle|\delta_J^{\mathrm{theory}}|\rangle=0.00730\%$ and $\langle|\delta_{\gamma_m}^{\mathrm{theory}}|\rangle=2.05\%$, respectively.

The theoretical limiting currents show close agreement with the independently obtained BVP estimates over the investigated parameter range. The mean absolute relative difference is approximately $0.0073\%$, which is smaller than the typical numerical uncertainty associated with the limiting-current search. A larger average difference is observed for the minimum Lorentz factor, indicating that agreement in the limiting current does not necessarily imply comparable accuracy in the internal potential minimum. Since both approaches are based on the same one-dimensional Poisson equation, further numerical refinement would be required to establish the origin of this difference. The resulting one-dimensional SCLC solution provides the longitudinal reference for the finite-width approximations developed in Sec.~III.

The limiting-current predictions exhibit excellent agreement with the BVP results, with a mean absolute relative difference below $0.01\%$ over the investigated parameter range. A larger average difference is observed for the minimum Lorentz factor, although the origin of this discrepancy has not been established. Since both approaches are based on the same one-dimensional Poisson equation, this difference may be associated with the numerical determination of the critical state near the turning point. The comparison provides a quantitative assessment of the stationary-current formulation, while the resulting longitudinal velocity and charge distributions form the basis for the finite-width approximations developed in Sec.~III.

\section{Two-Dimensional Relativistic Space-Charge-Limited Current}

We consider a planar diode with the cathode at $x=0$, the anode at $x=D$, and a centered emitting strip extending over $-W/2\leq y\leq W/2$. The geometry is invariant along $z$, and electrons are injected predominantly in the positive $x$ direction. The electrode potentials are $V_c$ and $V_a=V_c+V_g$, where $V_g>0$. The stationary relativistic 1D SCLC solution developed in Sec.~II provides the longitudinal velocity $v(x)$, the electron density, and the location $x_m$ of the potential minimum. We designate $0<x<x_m$ as region I and $x_m<x<D$ as region II. The position $x_m$ is determined by the 1D boundary conditions and limiting-current criterion and is retained in the subsequent 2D approximations.

We investigate two approaches to estimating the influence of finite emission width. The field-balance (FB) approach imposes a local longitudinal electric-field condition at $(x_m,0)$ and evaluates the field generated by the prescribed charge distribution using a free-space electrostatic kernel. The density-weighted (DW) approach uses a spatial moment of the same 1D distribution to construct a finite-width scaling estimate. Neither approach solves the coupled 2D Poisson and transport equations self-consistently. In particular, FB neglects the field induced by charges on the electrodes, whereas DW requires an additional assumption relating the density moment to the limiting current. Both retain the prescribed longitudinal distribution without accounting for transverse charge redistribution. Although longitudinal electron motion is treated relativistically, magnetic fields generated by the beam are neglected. Predictions at highly relativistic velocities should therefore be regarded as electrostatic approximations rather than complete electromagnetic descriptions.

\subsection{Field-Balance Approach}

We define the positive $x$ direction from the cathode toward the anode. Because $V_a>V_c$, the applied electric field points toward the cathode, with $E_x^{\rm vac}=-E_0$ and $E_0=(V_a-V_c)/D=V_g/D>0$. The quantity $\rho(x,y)>0$ denotes the magnitude of the electron charge density. At $(x_m,0)$, electrons in region I generate an electric field in the negative $x$ direction, whereas electrons in region II generate a field in the positive $x$ direction. With $E_I^{(2D)}$ and $E_{II}^{(2D)}$ denoting the respective positive field magnitudes, the local FB condition is
\begin{equation}
-E_0-E_I^{(2D)}+E_{II}^{(2D)}=0,
\qquad E_0+E_I^{(2D)}=E_{II}^{(2D)}.
\label{Eq_E_zero_2D}
\end{equation}
Using the free-space electrostatic kernel for a sheet of finite width, we express this condition as
\begin{equation}
\begin{aligned}
E_0&+\frac{1}{2\pi\varepsilon_0}\int_0^{x_m}\!\int_{-W/2}^{W/2}
\frac{\rho(x,y)(x_m-x)}{(x_m-x)^2+y^2}\,dy\,dx\\
&=\frac{1}{2\pi\varepsilon_0}\int_{x_m}^{D}\!\int_{-W/2}^{W/2}
\frac{\rho(x,y)(x-x_m)}{(x-x_m)^2+y^2}\,dy\,dx.
\end{aligned}
\label{Eq_E2d}
\end{equation}
The opposing field directions are already incorporated in Eq.~\eqref{Eq_E_zero_2D}, so both integrands are nonnegative. Assuming a uniform transverse charge distribution, $\rho(x,y)\simeq\rho(x)$, and retaining the 1D velocity profile, we write $\rho(x)=J_{\rm SCL}(2D)/v(x)$. Integration over the emitting width then yields
\begin{equation}
\begin{aligned}
E_0&+\frac{J_{\rm SCL}(2D)}{\pi\varepsilon_0}
\int_0^{x_m}\tan^{-1}\!\left[\frac{W}{2(x_m-x)}\right]\frac{dx}{v_I(x)}\\
&=\frac{J_{\rm SCL}(2D)}{\pi\varepsilon_0}
\int_{x_m}^{D}\tan^{-1}\!\left[\frac{W}{2(x-x_m)}\right]\frac{dx}{v_{II}(x)}.
\end{aligned}
\label{Eq_E2d_reduced}
\end{equation}
The inverse-tangent arguments are positive in both regions. An equivalent expression with region-I limits from $x_m$ to $0$ must retain the corresponding negative inverse-tangent argument. This distinction is relevant to the correction of the classical derivation in Appendix~\ref{App_2024_FB_correction}.

With the longitudinal velocity profile held fixed, the formal limit $W\to\infty$ of Eq.~\eqref{Eq_E2d_reduced} gives
\begin{equation}
E_0+\frac{J_{\rm SCL}(1D)}{2\varepsilon_0}
\int_0^{x_m}\frac{dx}{v_I(x)}
=\frac{J_{\rm SCL}(1D)}{2\varepsilon_0}
\int_{x_m}^{D}\frac{dx}{v_{II}(x)}.
\label{Eq_E1d_from_2d}
\end{equation}
The notation $J_{\rm SCL}(1D)$ is retained from Ref.~\cite{Liu2024scl} for the infinite-width FB reference. By comparison, the exact 1D Poisson solution subject to the electrode potentials satisfies
\begin{equation}
\frac{V_g}{D}=\frac{J_{\rm SCL}(1D)}{\varepsilon_0D}
\left[\int_{x_m}^{D}\frac{D-x}{v_{II}(x)}\,dx
-\int_0^{x_m}\frac{x}{v_I(x)}\,dx\right].
\label{Eq_FB_true_1D_balance}
\end{equation}
Equations~\eqref{Eq_E1d_from_2d} and \eqref{Eq_FB_true_1D_balance} are generally inequivalent because the free-space kernel omits the response of the conducting electrodes. Accordingly, the finite-width FB enhancement is calibrated to the independently determined 1D SCLC current from Sec.~II. This calibration does not make the FB approximation an exact solution of the 2D boundary-value problem.

Eliminating $E_0$ between Eqs.~\eqref{Eq_E2d_reduced} and \eqref{Eq_E1d_from_2d} and adopting this normalization, we obtain
\begin{widetext}
\begin{equation}
\frac{J_{\rm SCL}(2D)}{J_{\rm SCL}(1D)}
\simeq\left[1-\frac{2}{\pi}
\frac{\displaystyle\int_{x_m}^{D}\tan^{-1}\!\left[\frac{2(x-x_m)}{W}\right]\frac{dx}{v_{II}(x)}
-\displaystyle\int_0^{x_m}\tan^{-1}\!\left[\frac{2(x_m-x)}{W}\right]\frac{dx}{v_I(x)}}
{\displaystyle\int_{x_m}^{D}\frac{dx}{v_{II}(x)}-\displaystyle\int_0^{x_m}\frac{dx}{v_I(x)}}
\right]^{-1}.
\label{Eq_J2_J1}
\end{equation}
\end{widetext}
The subtraction of the two region contributions is required by their opposing electric-field directions. Expanding Eq.~\eqref{Eq_J2_J1} for $W\gg D$ gives
\begin{widetext}
\begin{equation}
\frac{J_{\rm SCL}(2D)}{J_{\rm SCL}(1D)}
\simeq1+\frac{4}{\pi W}
\frac{\displaystyle\int_{x_m}^{D}\frac{(x-x_m)\,dx}{v_{II}(x)}
-\displaystyle\int_0^{x_m}\frac{(x_m-x)\,dx}{v_I(x)}}
{\displaystyle\int_{x_m}^{D}\frac{dx}{v_{II}(x)}-\displaystyle\int_0^{x_m}\frac{dx}{v_I(x)}}.
\label{Eq_FB_moment_ratio}
\end{equation}
\end{widetext}

To evaluate Eq.~\eqref{Eq_J2_J1}, we substitute the relativistic 1D SCLC solution from Sec.~II. The longitudinal position relative to the potential minimum, obtained from Eqs.~\eqref{Eq_g_I} and \eqref{Eq_g_total}, is
\begin{equation}
x-x_m=
\begin{cases}
-\dfrac{Dg(\gamma;\gamma_m)}{g_I(\gamma_c,\gamma_m)+g_{II}(\gamma_a,\gamma_m)},&x<x_m,\\[3mm]
+\dfrac{Dg(\gamma;\gamma_m)}{g_I(\gamma_c,\gamma_m)+g_{II}(\gamma_a,\gamma_m)},&x>x_m.
\end{cases}
\label{Eq_x_xm_2D}
\end{equation}
The Lorentz factor decreases from $\gamma_c$ to $\gamma_m$ in region I and increases from $\gamma_m$ to $\gamma_a$ in region II. We introduce the nonnegative variable
\begin{equation}
p^2=\left(\frac{\gamma^2-1}{\gamma_m^2-1}\right)^{1/2}-1,
\label{Eq_p_definition}
\end{equation}
whose endpoint values are
\begin{equation}
\begin{aligned}
p_c^2&=\left(\frac{\gamma_c^2-1}{\gamma_m^2-1}\right)^{1/2}-1,\\
p_a^2&=\left(\frac{\gamma_a^2-1}{\gamma_m^2-1}\right)^{1/2}-1.
\end{aligned}
\label{Eq_p_boundaries}
\end{equation}
Thus, $p$ decreases from $p_c$ to zero in region I and increases from zero to $p_a$ in region II. The Lorentz factor and velocity can be written as
\begin{equation}
\begin{aligned}
\gamma(p)&=\sqrt{1+(\gamma_m^2-1)(1+p^2)^2},\\
v(p)&=\frac{c(1+p^2)\sqrt{\gamma_m^2-1}}{\sqrt{1+(\gamma_m^2-1)(1+p^2)^2}}.
\end{aligned}
\end{equation}
The position function defined in Eq.~\eqref{Eq_g_partial} takes the form
\begin{equation}
g(p)=\int_0^p
\frac{2(1+t^2)(\gamma_m^2-1)^{3/4}}{\sqrt{1+(\gamma_m^2-1)(1+t^2)^2}}\,dt.
\label{Eq_g_p}
\end{equation}
Equation~\eqref{Eq_x_xm_2D} then becomes
\begin{equation}
x-x_m=\begin{cases}
-\dfrac{Dg(p)}{g(p_c)+g(p_a)},&\text{region I},\\[3mm]
+\dfrac{Dg(p)}{g(p_c)+g(p_a)},&\text{region II}.
\end{cases}
\end{equation}
Differentiation of $g(p)$, together with the expression for $v(p)$, gives
\begin{equation}
\frac{1}{v(p)}\frac{dg(p)}{dp}=\frac{2(\gamma_m^2-1)^{1/4}}{c},
\end{equation}
The corresponding transit-time differential is therefore
\begin{equation}
\frac{dx}{v(x)}=\begin{cases}
-\dfrac{2D(\gamma_m^2-1)^{1/4}}{c[g(p_c)+g(p_a)]}\,dp,&\text{region I},\\[3mm]
+\dfrac{2D(\gamma_m^2-1)^{1/4}}{c[g(p_c)+g(p_a)]}\,dp,&\text{region II}.
\end{cases}
\label{Eq_dx_v_p}
\end{equation}
In region I, $dx/dp<0$ and the integration proceeds from $p_c$ to zero. The transit-time integrals in both regions are therefore positive and can be expressed as
\begin{equation}
\begin{aligned}
\int_0^{x_m}\frac{dx}{v_I(x)}&=\frac{2D(\gamma_m^2-1)^{1/4}}{c[g(p_c)+g(p_a)]}\int_0^{p_c}dp,\\
\int_{x_m}^{D}\frac{dx}{v_{II}(x)}&=\frac{2D(\gamma_m^2-1)^{1/4}}{c[g(p_c)+g(p_a)]}\int_0^{p_a}dp.
\end{aligned}
\label{Eq_FB_transit_p}
\end{equation}
Likewise, the inverse-tangent integrals become
\begin{widetext}
\begin{equation}
\begin{aligned}
\int_0^{x_m}\tan^{-1}\!\left[\frac{2(x_m-x)}{W}\right]\frac{dx}{v_I(x)}
&=\frac{2D(\gamma_m^2-1)^{1/4}}{c[g(p_c)+g(p_a)]}\int_0^{p_c}\tan^{-1}\!\left[\frac{2Dg(p)}{W[g(p_c)+g(p_a)]}\right]dp,\\[2mm]
\int_{x_m}^{D}\tan^{-1}\!\left[\frac{2(x-x_m)}{W}\right]\frac{dx}{v_{II}(x)}
&=\frac{2D(\gamma_m^2-1)^{1/4}}{c[g(p_c)+g(p_a)]}\int_0^{p_a}\tan^{-1}\!\left[\frac{2Dg(p)}{W[g(p_c)+g(p_a)]}\right]dp.
\end{aligned}
\label{Eq_FB_arctan_p}
\end{equation}
\end{widetext}
Substitution of Eqs.~\eqref{Eq_FB_transit_p} and \eqref{Eq_FB_arctan_p} into Eq.~\eqref{Eq_J2_J1} eliminates the common prefactors. Retaining the separate region contributions before combining the integrals gives
\begin{widetext}
\begin{equation}
\begin{aligned}
\frac{J_{\rm SCL}(2D)}{J_{\rm SCL}(1D)}
&\simeq\left[1-\frac{2}{\pi}
\frac{\displaystyle\int_0^{p_a}\tan^{-1}\!\left[\frac{2Dg(p)}{W[g(p_c)+g(p_a)]}\right]dp
-\displaystyle\int_0^{p_c}\tan^{-1}\!\left[\frac{2Dg(p)}{W[g(p_c)+g(p_a)]}\right]dp}{p_a-p_c}\right]^{-1}\\[2mm]
&=\left[1-\frac{2}{\pi(p_a-p_c)}\int_{p_c}^{p_a}
\tan^{-1}\!\left[\frac{2Dg(p)}{W[g(p_c)+g(p_a)]}\right]dp\right]^{-1}.
\end{aligned}
\label{Eq_J2_J1_relativistic}
\end{equation}
\end{widetext}
The initial expression retains the contributions of regions II and I separately. Their difference reduces to the single integral shown in the second expression, without altering the signed-field origin of the FB approximation.

For $W\gg D$, a first-order expansion of the inverse tangent and the outer reciprocal yields
\begin{equation}
\frac{J_{\rm SCL}(2D)}{J_{\rm SCL}(1D)}
\simeq1+\frac{4D}{\pi W}F_{\rm FB},
\label{Eq_J2_J1_wide}
\end{equation}
where
\begin{equation}
F_{\rm FB}=\frac{\displaystyle\int_{p_c}^{p_a}g(p)\,dp}
{[g(p_c)+g(p_a)](p_a-p_c)}.
\label{Eq_F_FB}
\end{equation}
The derivation assumes $\gamma_m>1$ and $V_g>0$, with the zero-injection limit obtained by continuity. The nonrelativistic correspondence and the singular nature of the symmetric drift-space limit are examined in Sec.~\ref{Sec_FB_DW_comparison} and Appendix~\ref{App_2024_FB_correction}.

\subsection{Density-Weighted Approach}

Unlike FB, the DW approach constructs a finite-width correction from a moment of the 1D charge distribution without imposing a local electric-field condition. Motivated by the classical scaling $1+FG$~\cite{Koh_2005}, we use the density-weighted absolute distance from the potential minimum to characterize the longitudinal distribution. Both regions contribute positively to this moment. The resulting DW correction is an independent scaling approximation rather than a rearrangement of Eq.~\eqref{Eq_J2_J1_relativistic}.

From Eq.~\eqref{Eq_x_xm_2D}, the normalized positions in regions I and II are
\begin{equation}
\begin{aligned}
\xi_I(p)
&=
\frac{x_m}{D}
-
\frac{g(p)}
{g_I(p_c)+g_{II}(p_a)},
\\
\xi_{II}(p)
&=
\frac{x_m}{D}
+
\frac{g(p)}
{g_I(p_c)+g_{II}(p_a)}.
\end{aligned}
\label{Eq_xi_p_2D}
\end{equation}
For either region, the normalized distance from the potential minimum is
\begin{equation}
\left|\xi-\frac{x_m}{D}\right|
=
\frac{g(p)}
{g_I(p_c)+g_{II}(p_a)}.
\label{Eq_distance_xm_p}
\end{equation}
We define the effective longitudinal factor as the density-weighted mean distance from $x_m$,
\begin{equation}
F_{\rm eff}
\equiv
\frac{
\displaystyle
\int_0^D
\left|\frac{x-x_m}{D}\right|
n(x)\,dx
}{
\displaystyle
\int_0^D n(x)\,dx
}.
\label{Eq_F_eff_definition}
\end{equation}
Equivalently, $F_{\rm eff}=\langle |x-x_m|\rangle/D$, which differs from the mean-position factor $F=\langle x\rangle/D$ introduced in Sec.~II. Both factors are evaluated from the same 1D SCLC distribution but refer to different spatial origins.

Separating the contributions on either side of $x_m$, Eq.~\eqref{Eq_F_eff_definition} becomes
\begin{equation}
F_{\rm eff}
=
\frac{
\displaystyle
\int_0^{x_m}
\frac{x_m-x}{D}n(x)\,dx
+
\int_{x_m}^{D}
\frac{x-x_m}{D}n(x)\,dx
}{
\displaystyle
\int_0^{x_m}n(x)\,dx
+
\int_{x_m}^{D}n(x)\,dx
}.
\label{Eq_F_eff_split}
\end{equation}
Because $n(x)=J/[ev(x)]$, the density weighting is proportional to the transit-time measure. Accounting for the opposite integration directions in the two regions, we use the positive measure $|dx|$ and write
\begin{equation}
n(x)\,|dx|
=
\frac{J}{e}\frac{|dx|}{v}.
\label{Eq_density_measure_p}
\end{equation}
From Eq.~\eqref{Eq_dx_v_p},
\begin{equation}
\frac{|dx|}{v}
=
\frac{2D}{c}
\frac{(\gamma_m^2-1)^{1/4}}
{g_I(p_c)+g_{II}(p_a)}
\,dp .
\label{Eq_dx_v_p_Feff}
\end{equation}
The common prefactor cancels between the numerator and denominator of Eq.~\eqref{Eq_F_eff_split}. With Eq.~\eqref{Eq_distance_xm_p}, the result is
\begin{equation}
F_{\rm eff}
=
\frac{1}{g_I(p_c)+g_{II}(p_a)}
\frac{
\displaystyle\int_0^{p_c}g(p)\,dp
+
\displaystyle\int_0^{p_a}g(p)\,dp
}{
p_a+p_c
}.
\label{Eq_F_eff_p}
\end{equation}
Using the geometrical scaling proposed in Ref.~\cite{Koh_2005}, we estimate the finite-width current enhancement as
\begin{equation}
\bar{J}_{\rm DW}
\equiv \frac{J_{\rm DW}}{J_{\rm SCL}^{(1D)}}
\simeq 1+\frac{4D}{\pi W}F_{\rm eff},
\qquad D/W\ll1.
\label{Eq_J2_J1_Feff}
\end{equation}
Equation~\eqref{Eq_J2_J1_Feff} combines the geometrical factor $4D/(\pi W)$ with the longitudinal factor $F_{\rm eff}$. This scaling relation is an approximation based on the 1D critical state and does not follow from the FB field integral.

The limiting cases provide useful consistency checks. When the potential minimum approaches the cathode, $x_m\rightarrow0$, the displacement satisfies $|x-x_m|\rightarrow x$, and thus
\begin{equation}
F_{\rm eff}\rightarrow F.
\label{Eq_Feff_to_F}
\end{equation}
In the nonrelativistic Child--Langmuir limit, $F\rightarrow1/4$, and Eq.~\eqref{Eq_J2_J1_Feff} reduces to the classical finite-width result in Eq.~\eqref{Eq_CL2d}~\cite{Luginsland_1996,Lau_2001}.

For zero injection in the weakly relativistic regime, $x_m=0$ and $F_{\rm eff}=F$, so the DW expression incorporates the correction in Eq.~\eqref{Eq_F_weak_rel} and approaches Eq.~\eqref{Eq_CL2d} as $U\to0$. 
At finite injection energy, the nonrelativistic limit of $F_{\rm eff}$ reproduces the classical factor $K(\epsilon)/4$ reported in Ref.~\cite{Liu2024scl}, as independently demonstrated in Appendix~\ref{App_2024_FB_correction}. Figure~\ref{Fig_F_F_eff} illustrates this correspondence.
Defining $s=\sqrt{\epsilon/(1+\epsilon)}$, the onrelativistic 1D profile gives
\begin{equation}
F_{\rm eff}^{\rm NR}=\frac{1+6s+6s^3+s^4}{4(1+s)^4}=\frac{K(\epsilon)}{4}.
\label{Eq_Feff_classical_exact}
\end{equation}
The equality holds throughout the classical finite-injection range. It reproduces the published coefficient algebraically but does not establish the DW scaling from the 2D Poisson equation. As $s\to1$, the classical drift-space limit is $F_{\rm eff}^{\rm NR}\to7/32$. The defining DW moment remains well defined in this symmetric limit.

Figure~\ref{Fig_F_F_eff} also illustrates the distinction between the cathode-referenced factor $F$ and the potential-minimum-referenced factor $F_{\rm eff}$. Only the latter enters the DW current estimate.

The relationship between $F$ and $F_{\rm eff}$ can be established by separating the charge distribution into regions I and II. We introduce
\begin{equation}
\begin{aligned}
N_I&=\int_0^{x_m}n(x)\,dx,\\
N_{II}&=\int_{x_m}^{D}n(x)\,dx,\\
N&=N_I+N_{II}.
\end{aligned}
\label{Eq_NI_NII}
\end{equation}
The first spatial moment in region I is
\begin{equation}
M_I=\int_0^{x_m}x\,n(x)\,dx,
\label{Eq_MI}
\end{equation}
From the definitions of $F$ and $F_{\rm eff}$, it follows that
\begin{equation}
F_{\rm eff}=F+\frac{x_m}{D}\frac{N_I-N_{II}}{N}-\frac{2M_I}{DN}.
\label{Eq_Feff_F_relation}
\end{equation}
Equation~\eqref{Eq_Feff_F_relation} shows that $F$ alone does not determine $F_{\rm eff}$, because the relative charge populations and the first moment in region I also contribute. The distinction is relevant to the DW construction, although the moment relation alone does not establish its accuracy as a 2D limiting-current estimate.

As the injection energy vanishes, $x_m\rightarrow0$ and $N_I,M_I\rightarrow0$, giving $F_{\rm eff}\rightarrow F$. At finite injection energy, the minimum lies inside the gap and the two factors generally differ. In particular, $F\rightarrow1/2$ for an increasingly uniform density distribution does not imply the same limit for $F_{\rm eff}$.

As shown in Fig.~\ref{Fig_F_F_eff}, $F_{\rm eff}$ also depends on $U$ through the longitudinal charge distribution. Agreement with $K(\epsilon)/4$ at small $U$ establishes a nonrelativistic benchmark but does not independently validate the finite-width DW scaling.

\begin{figure}[htbp]
    \centering
    \includegraphics[width=\columnwidth]{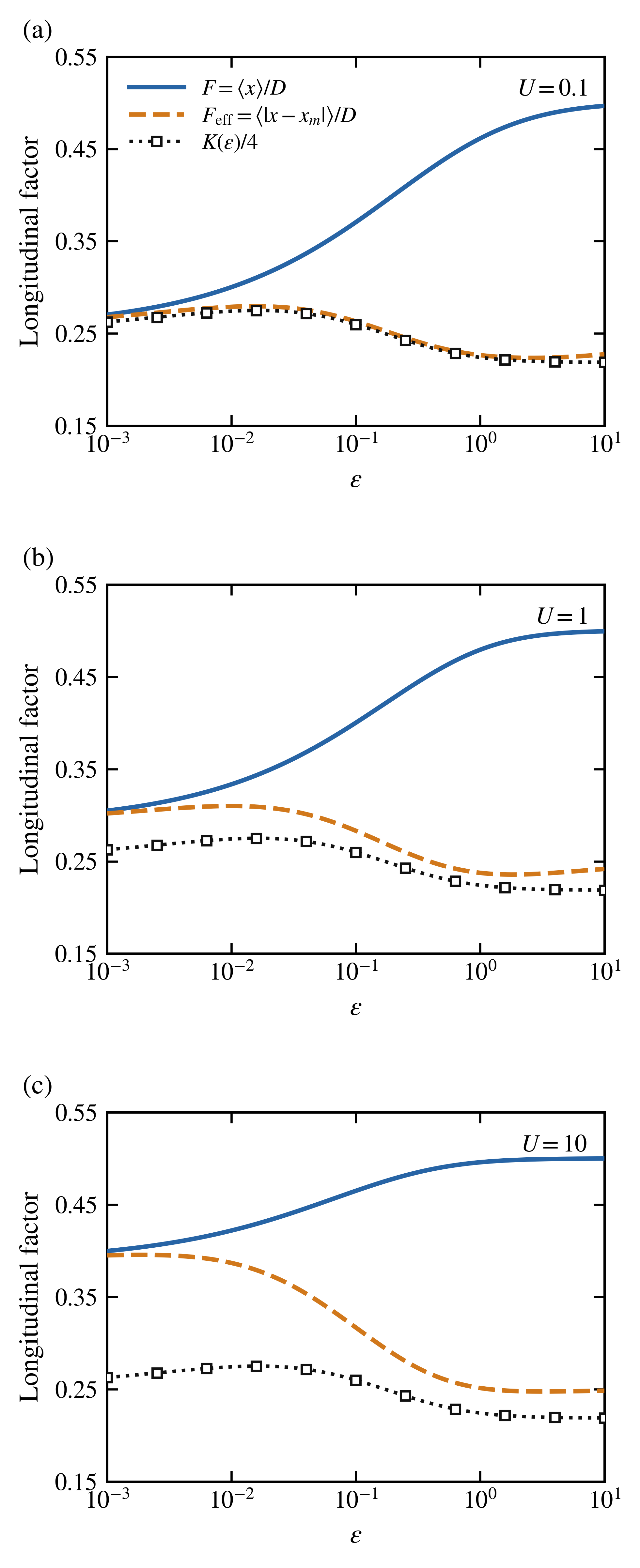}
    \caption{Relativistic longitudinal factors $F=\langle x\rangle/D$ and $F_{\rm eff}=\langle |x-x_m|\rangle/D$ as functions of $\epsilon$ for (a) $U=0.1$, (b) $U=1$, and (c) $U=10$. The former is referenced to the cathode and the latter to the potential minimum. The classical factor $K(\epsilon)/4$ is shown for comparison.
    }
    \label{Fig_F_F_eff}
\end{figure}

\subsection{Comparison of the Two Approaches}
\label{Sec_FB_DW_comparison}

The FB and DW approaches use the same longitudinal charge and velocity distributions but extract different quantities from them. FB determines the finite-width correction from the difference between oppositely directed space-charge fields, as reflected in Eqs.~\eqref{Eq_J2_J1} and \eqref{Eq_F_FB}. By contrast, DW uses the density-weighted absolute displacement from the potential minimum, defined in Eq.~\eqref{Eq_F_eff_definition}, to which both regions contribute positively. Replacing the subtractions in FB with additions would therefore change the underlying physical quantity and require an independent current-scaling assumption. It would not constitute a correction to the FB sign convention.

The resulting difference can be evaluated analytically for $D/W\ll1$. Combining Eqs.~\eqref{Eq_F_FB} and \eqref{Eq_F_eff_p} gives

\begin{equation}
F_{\rm FB}-F_{\rm eff}
=\frac{2\left[p_c\displaystyle\int_0^{p_a}g(p)\,dp-p_a\displaystyle\int_0^{p_c}g(p)\,dp\right]}
{[g(p_c)+g(p_a)](p_a^2-p_c^2)}.
\label{Eq_FB_DW_factor_difference}
\end{equation}

Since $g(p)$ is positive and strictly increasing, its average $p^{-1}\int_0^p g(t)\,dt$ also increases with $p$. Therefore, $F_{\rm FB}>F_{\rm eff}$ for $p_a>p_c>0$, whereas the factors coincide as $p_c\to0$. FB consequently predicts the larger leading-order enhancement at finite injection energy. This ordering follows from the respective definitions and does not establish which prediction is more accurate.

The nonrelativistic limit provides a direct comparison with earlier work. Appendix~\ref{App_2024_FB_correction} shows that relativistic FB reduces to the sign-corrected coefficient $K_{\rm FB}(\epsilon)$ in Eq.~\eqref{Eq_App2024_K_FB}, whereas DW yields $K_{\rm DW}(\epsilon)$ in Eq.~\eqref{Eq_App2024_K_DW}. The DW coefficient agrees with $K(\epsilon)$ reported in Eq.~(29) of Ref.~\cite{Liu2024scl}, but the corrected FB coefficient does not. The discrepancy originates from the treatment of the region-I integration direction between Eqs.~(20) and (21) of that reference and the inverse-tangent branch used in Eq.~(23). As shown in Eq.~\eqref{Eq_App2024_K_difference}, the difference between the coefficients is positive for $\epsilon>0$, approaches $2\sqrt{\epsilon}$ for $\epsilon\ll1$, and tends to $9/8$ in the formal classical drift-space limit. These results identify distinct classical limits for the two relativistic constructions.

The drift-space limit further distinguishes the two approaches. For $V_g\to0^+$ at finite classical injection energy, $K_{\rm FB}\to2$ and $K_{\rm DW}\to7/8$, corresponding to $F_{\rm FB}\to1/2$ and $F_{\rm eff}\to7/32$. The DW result reproduces the classical first-order coefficient obtained in Ref.~\cite{Liu_2014}, whereas the present two-sided FB construction does not. This discrepancy does not arise from an incorrect evaluation of Eq.~\eqref{Eq_F_FB}. For $p_a>p_c$, the FB factor is
\begin{equation}
F_{\rm FB}=\frac{\displaystyle\int_{p_c}^{p_a}g(p)\,dp}
{[g(p_c)+g(p_a)](p_a-p_c)},
\label{Eq_FB_mean_value_limit}
\end{equation}
which approaches $1/2$ as $p_a\to p_c>0$. This mathematical limit remains finite even though the underlying FB condition becomes degenerate at $V_g=0$. At exact symmetry, the opposing space-charge fields cancel, and $E_I^{(2D)}=E_{II}^{(2D)}$ no longer imposes an independent constraint on the current. The finite limit therefore results from the normalization of two vanishing field differences rather than a limiting-current condition for the symmetric diode.

The same distinction appears in the full inverse-tangent expression. Taking $p_a\to p_c>0$ in Eq.~\eqref{Eq_J2_J1_relativistic} gives
\begin{equation}
\lim_{p_a\to p_c}\bar J_{\rm FB}
=\left[1-\frac{2}{\pi}\tan^{-1}\!\left(\frac{D}{W}\right)\right]^{-1}.
\label{Eq_FB_formal_drift}
\end{equation}
The formal limit is independent of the longitudinal velocity distribution, reflecting the cancellation inherent in the symmetric field balance rather than a physical current threshold. In contrast, the drift-space calculation of Ref.~\cite{Liu_2014} compared nonzero, one-sided space-charge fields in the finite-width and infinite-width geometries. Its symmetric half-gap expression is
\begin{equation}
\frac{J_{\rm SCL}(2D)}{J_{\rm SCL}(1D)}
=\left[1-\frac{2}{\pi}
\frac{\displaystyle\int_{D/2}^{D}\tan^{-1}\!\left[\frac{2(x-D/2)}{W}\right]\frac{dx}{v(x)}}
{\displaystyle\int_{D/2}^{D}\frac{dx}{v(x)}}\right]^{-1}.
\label{Eq_FB_2014_one_sided}
\end{equation}
The one-sided reference remains nonzero, preserving the velocity-dependent weighting at $V_g=0$. The two methods thus impose different field comparisons. The 2014 expression cannot be recovered by directly setting $V_g=0$ in the present two-sided FB condition, and neither approximation independently determines a self-consistent 2D limiting current.

\subsection{Comparison with Particle-in-Cell Simulations}
\begin{figure*}[t]
\centering
\includegraphics[width=\textwidth]{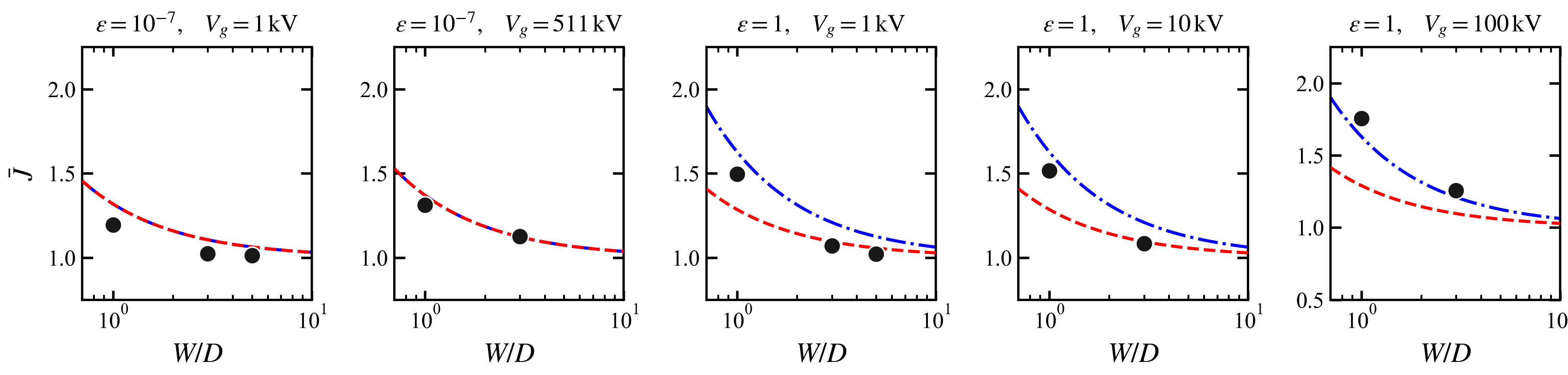}
\caption{Normalized finite-width space-charge-limited current, $\bar J=J_{\rm SCL}(2D)/J_{\rm SCL}(1D)$, obtained from XOOPIC electrostatic PIC simulations (black circles), the leading-order FB model (blue dash-dotted lines), and the DW model (red dashed lines). The analytical predictions follow Eqs.~\eqref{Eq_J2_J1_wide} and \eqref{Eq_J2_J1_Feff} and use the same independently determined 1D reference current. The emitter width is normalized by the gap spacing $D$.}
\label{Fig_PIC}
\end{figure*}

We evaluate the finite-width estimates using two-dimensional electrostatic (ES) particle-in-cell (PIC) simulations with XOOPIC~\cite{Verboncoeur1995}. The diode has a centered emitting strip with fixed gap spacing $D=0.01~\mathrm{m}$ and varying emitter width $W$. The applied voltages $V_g=1$, $10$, $100$, and $511~\mathrm{kV}$ correspond to $U=eV_g/(m_ec^2)\simeq0.00196$, $0.0196$, $0.196$, and $1$, respectively. We consider nearly zero injection energy ($\epsilon=10^{-7}$) and finite injection energy ($\epsilon=1$). Because ES simulations omit the magnetic field generated by the electron beam, they provide a comparison within electrostatic transport rather than a test of fully electromagnetic relativistic dynamics.

The XOOPIC coordinates differ from those used in the analytical derivation. The longitudinal direction is $y$, with electrodes at $y=0$ and $y=D$, whereas the transverse direction is $x$, with $0\leq x\leq L_x$ and $L_x=10W$. The emitting strip is centered on the cathode, and the lateral boundaries are conducting. The grid spacings are $\Delta x=10^{-3}~\mathrm{m}$ and $\Delta y=10^{-4}~\mathrm{m}$. Electrons are injected with kinetic energy $K_0=\epsilon eV_g$, Lorentz factor $\gamma_0=1+\epsilon U$, and speed $v_0=c\sqrt{1-\gamma_0^{-2}}$. The time step is $\Delta t=1.25\times10^{-12}~\mathrm{s}$, with 32 macroparticles injected per emitting cell at each step. The injection current is varied to determine the onset of a virtual-cathode barrier, beyond which the full injected current can no longer be transmitted steadily. For finite injection energy, the barrier threshold corresponds to the loss of sufficient longitudinal kinetic energy for transmission.

The simulated current enhancement is normalized as
\begin{equation}
\bar J_{\rm PIC}=\frac{J_{\rm SCL}^{(2D,\rm PIC)}}{J_{\rm SCL}^{(1D)}},
\label{Eq_J_PIC_normalization}
\end{equation}
where $J_{\rm SCL}^{(1D)}$ is the independently determined relativistic 1D limiting current. The FB and DW curves in Fig.~\ref{Fig_PIC} use Eqs.~\eqref{Eq_J2_J1_wide} and \eqref{Eq_J2_J1_Feff}, respectively, so both analytical estimates are evaluated to the same leading order in $D/W$.

At $\epsilon=10^{-7}$, FB and DW give nearly identical predictions because region I contributes negligibly and $F_{\rm FB}\simeq F_{\rm eff}$. Agreement between the analytical models does not, however, ensure agreement with PIC. For $V_g=1~\mathrm{kV}$ and $W/D=1$, both predict approximately $1.319$, compared with the PIC value of $1.1955$. At $V_g=511~\mathrm{kV}$ and $W/D=3$, their common prediction of approximately $1.123$ is close to the PIC result of $1.1254$. These cases illustrate the variation in accuracy with diode parameters without identifying a unique source of error.

At $\epsilon=1$, FB predicts larger enhancements than DW, consistent with Eq.~\eqref{Eq_FB_DW_factor_difference}. For $V_g=1~\mathrm{kV}$ and $W/D=1$, the FB and DW predictions are $1.6273$ and $1.2854$, respectively, while PIC gives $1.4957$. At $W/D=3$, the corresponding values are $1.2091$ and $1.0951$, compared with $1.0705$ from PIC, making DW the closer estimate. At $V_g=100~\mathrm{kV}$, PIC gives $1.7554$ and $1.2566$ for $W/D=1$ and $3$, respectively. Both values exceed the analytical predictions, although FB ($1.6305$ and $1.2102$) is closer than DW ($1.2909$ and $1.0970$). Neither approach is therefore uniformly more accurate across the simulated cases.

For $\epsilon=1$ and $V_g=511~\mathrm{kV}$, two additional PIC cases not included in Fig.~\ref{Fig_PIC} give $\bar J_{\rm PIC}=2.8998$ and $2.0866$ at $W/D=1$ and $3$, respectively. Both values substantially exceed the FB predictions ($1.6347$ and $1.2116$) and the DW predictions ($1.3024$ and $1.1008$). Here $\gamma_c=2$ and $\gamma_a=3$, so magnetic self-fields may be important in a physical electromagnetic diode. However, these fields are absent from both the analytical models and the ES simulations. The observed discrepancies must therefore arise within the electrostatic comparison and cannot be attributed to omitted magnetic forces alone.

The comparison is also affected by the approximations in the analytical models. PIC permits the charge distribution and potential to respond to emitting-edge fields and conducting boundaries, whereas FB and DW prescribe the longitudinal 1D profile. FB additionally omits the electrode response in its free-space kernel. Furthermore, the first-order formulas are valid asymptotically for $D/W\ll1$ and are extrapolated when $W/D\sim1$. Consequently, the observed discrepancies cannot be attributed solely to the choice of longitudinal factor, nor do they establish a general ranking of FB and DW.

\section{Discussion}

The present work develops a relativistic 1D SCLC description for arbitrary injection energy and uses its stationary solution to construct two finite-width estimates. The 1D theory satisfies the longitudinal Poisson and energy equations together with the electrode potentials and the limiting-current condition. By contrast, FB and DW retain the 1D critical profile without solving for transverse charge redistribution. Their agreement with known classical coefficients provides useful consistency checks but does not establish the validity of a self-consistent 2D SCLC theory.

FB retains an explicit description of the signed electrostatic contributions. With the applied field directed toward the cathode, the fields generated by regions I and II oppose one another at the potential minimum. Their directions determine the local condition $-E_0-E_I+E_{II}=0$. In Ref.~\cite{Liu2024scl}, the region-I contribution was correctly oriented in the finite-width integral, but its sign was not preserved consistently in the subsequent infinite-width reduction. Appendix~\ref{App_2024_FB_correction} examines the published Eqs.~(20) through (24) and obtains the corrected classical coefficient in Eq.~\eqref{Eq_App2024_K_FB}. The present relativistic FB result reduces to this sign-consistent coefficient as $U\to0$ at fixed finite $\epsilon$, but not to the different $K(\epsilon)$ reported in Eq.~(29) of Ref.~\cite{Liu2024scl}. The distinction is necessary when identifying which classical construction the relativistic theory recovers.

A further limitation concerns the infinite-width reference used by FB. The free-space Coulomb kernel yields an explicit charge-generated field but does not incorporate the response of charges induced on the electrodes. Consequently, its $W\to\infty$ field-balance relation need not reproduce the exact 1D Poisson limiting current. The boundary-consistent equation, Eq.~\eqref{Eq_FB_true_1D_balance}, contains different longitudinal position weights. For the classical zero-injection Child--Langmuir solution, for example, the free-space balance evaluated with the exact 1D current yields $2E_0/3$ rather than $E_0$. We therefore calibrate the FB enhancement to the independently determined 1D current. This choice preserves the expected normalized infinite-width limit but does not restore the missing electrode fields or establish a self-consistent 2D limiting-current condition. In particular, a local zero-field condition at an assumed potential minimum is not, by itself, sufficient to determine the maximum steady transmitted current.

This limitation becomes particularly apparent in the drift-space limit. At $V_g=0$, the symmetric left and right contributions cancel, so the two-sided FB condition holds independently of the current amplitude. The corresponding normalized expression involves a ratio of two vanishing field differences. Approaching symmetry from $V_g>0$ gives the finite formal limit in Eq.~\eqref{Eq_FB_formal_drift}, with $K_{\rm FB}\to2$ in the classical wide-emitter regime. This limit is mathematically consistent but does not determine a threshold current at exact symmetry. By contrast, Ref.~\cite{Liu_2014} compared nonzero one-sided fields and retained the velocity-dependent half-gap integral, as shown in Eq.~\eqref{Eq_FB_2014_one_sided}. The different constraints explain why the earlier drift-space expression cannot be obtained by directly setting $V_g=0$ in the present FB formulation. Neither construction, however, satisfies the full conducting-boundary problem independently.

The DW factor remains well defined in the symmetric limit because $F_{\rm eff}=\langle|x-x_m|\rangle/D$ is a positive displacement moment. It reduces to $F=\langle x\rangle/D$ as $x_m\to0$, and its nonrelativistic coefficient in Eq.~\eqref{Eq_App2024_K_DW} agrees with the published $K(\epsilon)$, as demonstrated in Appendix~\ref{App_2024_FB_correction}. In the classical drift-space limit, $F_{\rm eff}\to7/32$, reproducing the corresponding first-order coefficient $7/8$. These limits support DW as a consistent classical interpolation across injection regimes. Nevertheless, the connection between the displacement moment and the 2D limiting current is assumed rather than derived from the coupled Poisson and transport equations. Distinct transverse charge distributions can share the same longitudinal moment while producing different electric fields, so agreement with classical limiting coefficients does not establish general predictive validity.

The analytic ordering $F_{\rm FB}>F_{\rm eff}$ at positive injection energy follows from Eq.~\eqref{Eq_FB_DW_factor_difference}. In the classical regime, their coefficient difference increases toward $9/8$ in the formal drift-space limit, as shown in Eq.~\eqref{Eq_App2024_K_difference}. This difference reflects the normalization of a signed field difference in FB and a positive density moment in DW. It does not represent an additional physical field or provide a measure of relative accuracy. Electrode response and transverse redistribution can affect the actual 2D threshold independently of this ordering.

The PIC comparison confirms that neither FB nor DW provides a uniformly accurate prediction of the 2D limiting current. Although the simulated values fall between the two estimates in some cases, this behavior is not universal, and the two approaches do not establish upper and lower bounds on the SCLC. At finite injection energy, the PIC results may fall below or exceed both predictions, depending on the applied voltage and emitter width. In particular, both models substantially underestimate the highest-voltage ES-PIC results. These discrepancies cannot be attributed to magnetic forces, which are absent from both the analytical models and the simulations. Instead, the comparisons highlight the limitations of the prescribed 1D charge distributions and approximate finite-width treatments. The possible influence of magnetic self-fields in a fully electromagnetic diode remains outside the scope of the present study.

Both approaches prescribe the 1D longitudinal profile despite the transverse electric fields and charge redistribution associated with a finite emitter. The potential minimum and density profile of an actual 2D threshold state may therefore differ from those used in the approximations. Furthermore, the first-order expansions assume $W\gg D$, whereas some simulated cases have $W/D=1$. Discrepancies in this range cannot be attributed uniquely to the longitudinal correction factors. Further assessment would benefit from wider-emitter scans and diagnostics of the transverse potential and density distributions near the simulated threshold.

A more complete electrostatic treatment would use a Green function that satisfies the conducting-electrode boundary conditions. With signed charge density $\rho_s(\mathbf r)$ and Dirichlet Green function $G_D$, the potential can be expressed as
\begin{equation}
\phi(\mathbf r)=\phi_{\rm vac}(\mathbf r)+\frac{1}{\varepsilon_0}\int_{\Omega}G_D(\mathbf r,\mathbf r')\rho_s(\mathbf r')\,d^2\mathbf r',
\label{Eq_Green_future}
\end{equation}
where $-\nabla^2G_D(\mathbf r,\mathbf r')=\delta(\mathbf r-\mathbf r')$ and $G_D$ vanishes at the conducting boundaries. This formulation accounts for the electrode response. However, replacing the free-space kernel alone would not make the calculation self-consistent if the charge distribution remained fixed in its 1D form. The potential, density, transverse structure, position of the minimum, and limiting-current condition must instead be determined together. Lau et al.~\cite{Lau2026} have developed complementary differential, integral, and PIC approaches for nonrelativistic diodes with small emitting areas and low injection energy. Their boundary-consistent methodology offers a useful starting point, although their specific narrow-emitter scaling laws do not directly apply to the relativistic parameter range considered here.

Overall, FB provides an explicit signed-field formulation for asymmetric diodes but requires calibration to the physical 1D reference and becomes degenerate at exact drift-space symmetry. DW gives a regular displacement moment that reproduces the published classical finite-injection and drift-space coefficients, although its current scaling remains an assumption. Both are leading-order finite-width estimates whose applicability depends on voltage, injection energy, and emitter width. A predictive 2D SCLC theory requires a self-consistent treatment of charge transport and the conducting-electrode boundary conditions.

\section{Conclusion}

We have developed a relativistic one-dimensional SCLC formulation for arbitrary injection energy, connecting the accelerating-diode and drift-space regimes through a common stationary limiting-current condition. The resulting current predictions agree closely with independent numerical boundary-value solutions, with a mean absolute relative difference of approximately 0.0073\% over the investigated parameter range. The corresponding longitudinal charge distribution provides the reference for evaluating finite-width effects.

Using this 1D weakly relativistic SCLC model,  we have examined two different models to account for the effects of finite emission area. The FB approach evaluates the opposing space-charge fields at the potential minimum, whereas the DW approach uses the density-weighted mean absolute displacement from that minimum. The two approaches give different corrections at finite injection energy, even in the non-relativistic limit, with FB predicting a larger leading-order enhancement. DW reproduces the previously published classical finite-injection coefficient and the classical drift-space coefficient of 7/8, whereas the corrected FB formulation approaches a coefficient of 2 in the formal classical drift-space limit. These differences reflect their distinct assumptions and do not establish the relative accuracy of the two approaches.

Comparisons with XOOPIC electrostatic PIC simulations show that neither approximation is uniformly more accurate, and their predictions do not constitute general upper and lower bounds on the 2D limiting current. 
The agreement depends on injection energy, applied voltage, and emitter width. Both approaches retain a prescribed 1D charge distribution and therefore neglect self-consistent transverse charge redistribution, while FB additionally omits the conducting-electrode response in its free-space field calculation. The present results clarify the classical correspondence and limitations of the two approximations and provide a basis for more self-consistent multidimensional treatments as shown in a recent paper even the simplest classical regime \cite{Lau2026}.

\section*{Acknowledgments}
This work was supported by the National Science and Technology Council (NSTC), Taiwan, under Grant No. NSTC 114-2112-M-006-021-MY3. The SUTD authors acknowledge support from the Innovative Undergraduate Research Programme (IROP).

\appendix
\section{Classical Comparison of the Field-Balance and Density-Weighted Approaches}
\label{App_2024_FB_correction}

We compare the classical limits of the field-balance (FB) and density-weighted (DW) approaches. We first re-examine the FB derivation in Ref.~\cite{Liu2024scl}, identifying inconsistencies in the integration direction and inverse-tangent branch that affect the published finite-width coefficient. We then derive the corrected FB coefficient and show that the previously published coefficient is independently recovered from the DW formulation. This comparison clarifies the distinct assumptions and classical limits of the two approaches.

\subsection{Correction of the field-balance integrals}

The positive $x$ direction is taken from the cathode at $x=0$ to the anode at $x=D$. For $V_a>V_c$, the applied electric field has a signed component $-E_0$, where $E_0=(V_a-V_c)/D>0$. At the potential minimum $x_m$, the space-charge fields generated by electrons in regions I and II point in the negative and positive $x$ directions, respectively. Defining $E_I$ and $E_{II}$ as their positive magnitudes, the field-balance condition is $-E_0-E_I+E_{II}=0$, or equivalently $E_0+E_I=E_{II}$. Equation~(20) of Ref.~\cite{Liu2024scl} can be written as
\begin{equation}
\begin{aligned}
E_0&+\frac{J_{\rm SCL}(2D)}{\pi\varepsilon_0}
\int_{x_m}^{0}\tan^{-1}\!\left[\frac{W}{2(x-x_m)}\right]\frac{dx}{v_I(x)}\\
&=\frac{J_{\rm SCL}(2D)}{\pi\varepsilon_0}
\int_{x_m}^{D}\tan^{-1}\!\left[\frac{W}{2(x-x_m)}\right]\frac{dx}{v_{II}(x)}.
\end{aligned}
\label{Eq_App2024_eq20}
\end{equation}
The region-I contribution is positive because both the inverse-tangent argument and the integration direction are negative. In particular,
\begin{equation}
\begin{aligned}
&\int_{x_m}^{0}\tan^{-1}\!\left[\frac{W}{2(x-x_m)}\right]\frac{dx}{v_I(x)}\\
&\quad=\int_0^{x_m}\tan^{-1}\!\left[\frac{W}{2(x_m-x)}\right]\frac{dx}{v_I(x)}.
\end{aligned}
\label{Eq_App2024_reverse}
\end{equation}
Thus, the finite-width field integral in the published Eq.~(20) is correctly oriented. Its infinite-width limit must instead be written as
\begin{equation}
\begin{aligned}
E_0&+\frac{J_{\rm SCL}(1D)}{2\varepsilon_0}
\int_0^{x_m}\frac{dx}{v_I(x)}\\
&=\frac{J_{\rm SCL}(1D)}{2\varepsilon_0}
\int_{x_m}^{D}\frac{dx}{v_{II}(x)}.
\end{aligned}
\label{Eq_App2024_eq21_correct}
\end{equation}
In contrast, the published Eq.~(21) retains the reversed region-I limits while assigning a positive sign to that term. This is inconsistent with the finite-width expression. The large-$W$ expansion also requires the appropriate inverse-tangent branch in each region,
\begin{equation}
\tan^{-1}\!\left[\frac{W}{2(x-x_m)}\right]\simeq
\begin{cases}
-\dfrac{\pi}{2}-\dfrac{2(x-x_m)}{W},&x<x_m,\\[2mm]
+\dfrac{\pi}{2}-\dfrac{2(x-x_m)}{W},&x>x_m.
\end{cases}
\label{Eq_App2024_arctan_correct}
\end{equation}
The positive $\pi/2$ branch used in Eq.~(23) of Ref.~\cite{Liu2024scl} applies only in region II. Retaining the correct branch in region I and eliminating $E_0$ yields the corrected counterpart of Eq.~(24),
\begin{equation}
\begin{aligned}
\frac{J_{\rm SCL}(2D)}{J_{\rm SCL}(1D)}&\simeq \\
&\left[1-\frac{4}{\pi W}
\frac{\displaystyle\int_{x_m}^{D}\frac{(x-x_m)\,dx}{v_{II}(x)}-\displaystyle\int_0^{x_m}\frac{(x_m-x)\,dx}{v_I(x)}}
{\displaystyle\int_{x_m}^{D}\frac{dx}{v_{II}(x)}-\displaystyle\int_0^{x_m}\frac{dx}{v_I(x)}}\right]^{-1}.
\label{Eq_App2024_FB_correct24}
\end{aligned}
\end{equation}
A literal evaluation of the reversed region-I integral in the published Eq.~(24) instead produces a sum of positive transit times in the denominator. Such a result does not follow from the signed-field balance. As discussed in the main text, the infinite-width reference associated with this free-space approximation is calibrated to the independently obtained one-dimensional SCLC current. The normalization should not be interpreted as an exact consequence of the one-dimensional Poisson boundary-value problem.

Evaluating Eq.~\eqref{Eq_App2024_FB_correct24} with the classical one-dimensional solution gives
\begin{align}
K_{\rm FB}(\epsilon)
&=1+\frac{4\sqrt{\dfrac{\epsilon}{1+\epsilon}}}
{\left(1+\sqrt{\dfrac{\epsilon}{1+\epsilon}}\right)^2}
\nonumber\\
&\simeq
\begin{cases}
1+4\sqrt{\epsilon},&\epsilon\ll1,\\[2mm]
2-\dfrac{1}{16\epsilon^2},&\epsilon\gg1.
\end{cases}
\label{Eq_App2024_K_FB}
\end{align}
The corresponding first-order finite-width current enhancement is
\begin{equation}
\frac{J_{\rm SCL}(2D)}{J_{\rm SCL}(1D)}
\simeq1+\frac{D}{\pi W}K_{\rm FB}(\epsilon).
\label{Eq_App2024_FB_K}
\end{equation}
This coefficient differs from $K(\epsilon)$ reported in Eq.~(29) of Ref.~\cite{Liu2024scl}, but agrees with the nonrelativistic limit of the present relativistic FB formulation, Eq.~\eqref{Eq_F_FB}.

\subsection{Classical density-weighted correction}

We next evaluate the DW correction independently of the electric-field balance, using $F_{\rm eff}$ defined in Eq.~\eqref{Eq_F_eff_definition}. In the classical stationary solution, the electron density is proportional to $1/v(x)$. The density-weighted mean distance from the potential minimum can therefore be expressed as
\begin{equation}
\begin{aligned}
F_{\rm eff}
&=\frac{1}{D}
\left(\int_0^{x_m}\frac{dx}{v_I(x)}
+\int_{x_m}^{D}\frac{dx}{v_{II}(x)}\right)^{-1}\\
&\quad\times\left(
\int_0^{x_m}\frac{(x_m-x)\,dx}{v_I(x)}
+\int_{x_m}^{D}\frac{(x-x_m)\,dx}{v_{II}(x)}
\right)\\
&=\frac{K_{\rm DW}(\epsilon)}{4}.
\end{aligned}
\label{Eq_App2024_DW_factor}
\end{equation}
Evaluation with the same classical one-dimensional solution yields
\begin{align}
K_{\rm DW}(\epsilon)
&=1+\frac{2\sqrt{\dfrac{\epsilon}{1+\epsilon}}}
{\left(1+\sqrt{\dfrac{\epsilon}{1+\epsilon}}\right)^2}
-10\left[
\frac{\sqrt{\dfrac{\epsilon}{1+\epsilon}}}
{\left(1+\sqrt{\dfrac{\epsilon}{1+\epsilon}}\right)^2}
\right]^2\nonumber\\
&\simeq
\begin{cases}
1+2\sqrt{\epsilon},&\epsilon\ll1,\\[2mm]
\dfrac78+\dfrac{3}{64\epsilon^2},&\epsilon\gg1.
\end{cases}
\label{Eq_App2024_K_DW}
\end{align}
This coefficient is algebraically identical to $K(\epsilon)$ published in Eq.~(29) of Ref.~\cite{Liu2024scl}. Substituting it into the DW scaling relation, Eq.~\eqref{Eq_J2_J1_Feff}, gives
\begin{equation}
\frac{J_{\rm SCL}(2D)}{J_{\rm SCL}(1D)}
\simeq1+\frac{D}{\pi W}K_{\rm DW}(\epsilon).
\label{Eq_App2024_DW_K}
\end{equation}
Thus, DW reproduces the published classical finite-injection coefficient at first order in $D/W$. The higher-order terms implied by the reciprocal form of Eq.~(28) in Ref.~\cite{Liu2024scl} do not follow from the present first-order DW scaling.

\subsection{Comparison of the two classical corrections}

The FB and DW approaches yield different correction factors despite using the same classical one-dimensional SCLC solution. Their difference is
\begin{equation}
\begin{aligned}
&K_{\rm FB}(\epsilon)-K_{\rm DW}(\epsilon)\\
&=\frac{2\sqrt{\dfrac{\epsilon}{1+\epsilon}}}
{\left(1+\sqrt{\dfrac{\epsilon}{1+\epsilon}}\right)^2}+10\left[
\frac{\sqrt{\dfrac{\epsilon}{1+\epsilon}}}
{\left(1+\sqrt{\dfrac{\epsilon}{1+\epsilon}}\right)^2}
\right]^2\\
&\simeq
\begin{cases}
2\sqrt{\epsilon},&\epsilon\ll1,\\[2mm]
\dfrac98-\dfrac{7}{64\epsilon^2},&\epsilon\gg1.
\end{cases}
\end{aligned}
\label{Eq_App2024_K_difference}
\end{equation}
The difference is positive for every finite $\epsilon>0$, so FB predicts a larger leading-order finite-width enhancement than DW. This difference reflects their distinct physical constructions. FB normalizes the difference between oppositely directed electric fields by the corresponding infinite-width field difference, whereas DW uses a positive density-weighted displacement for which the contributions from both regions are added. Their disagreement therefore does not represent an additional physical contribution or an alternative sign convention for the same field integral.

Both approaches recover the classical zero-injection Child--Langmuir coefficient. In the formal drift-space limit, however, $K_{\rm FB}\to2$ and $K_{\rm DW}\to7/8$. The latter agrees with the first-order classical coefficient obtained in Ref.~\cite{Liu_2014}. At exactly $V_g=0$, the two opposing fields in the symmetric FB configuration cancel, so the two-sided balance no longer supplies an independent limiting-current condition. The earlier drift-space formulation instead compared nonvanishing one-sided fields. Consequently, the present relativistic FB model reduces to the corrected classical FB result, whereas its DW counterpart reproduces the finite-injection coefficient published in Ref.~\cite{Liu2024scl}. Neither approximation constitutes a self-consistent solution of the two-dimensional Poisson problem with conducting-electrode boundary conditions.

\nocite{ref_SCL}
\bibliography{ref_SCL}

\end{document}